\documentstyle[11pt,aaspp4,epsf]{article}

\newcommand{\etal}{et al.\ }
\newcommand{\Hm}{{\cal F}}
\newcommand{\beq}{\begin{eqnarray}}
\newcommand{\feq}{\end{eqnarray}}
\def\go{\mathrel{\raise.3ex\hbox{$>$}\mkern-14mu\lower0.6ex\hbox{$\sim$}}}
\def\lo{\mathrel{\raise.3ex\hbox{$<$}\mkern-14mu\lower0.6ex\hbox{$\sim$}}}

\lefthead{Ford, Kozinsky, \& Rasio}
\righthead{Secular Evolution of Triples}

\begin{document}

\title{Secular Evolution of Hierarchical Triple Star Systems}
\date{now}
\author{Eric B.~Ford\altaffilmark{1}, Boris Kozinsky\altaffilmark{2},  
        \& Frederic A.~Rasio\altaffilmark{3}}
\affil{Physics Department, Massachusetts Institute of Technology 
  Cambridge, MA 02139 }

\altaffiltext{1}{Present address: Princeton University, Department of
Astrophysical Sciences, Peyton Hall, NJ 08544; eford@princeton.edu}
\altaffiltext{2}{bkoz@mit.edu}
\altaffiltext{3}{Alfred P.\ Sloan Research Fellow; rasio@mit.edu}

\begin{abstract}
 
We derive octupole-level secular perturbation equations for
hierarchical triple systems, using classical Hamiltonian perturbation
techniques.  Our equations describe the secular evolution of the
orbital eccentricities and inclinations over timescales long compared
to the orbital periods. By extending previous work done to leading
(quadrupole) order to octupole level (i.e., including terms of order
$\alpha^3$, where $\alpha\equiv a_1/a_2<1$ is the ratio of semimajor
axes) we obtain expressions that are applicable to a much wider range
of parameters.  In particular, our results can be applied to
high-inclination as well as coplanar systems, and our expressions are
valid for almost all mass ratios for which the system is in a stable
hierarchical configuration.  In contrast, the standard
quadrupole-level theory of Kozai gives a vanishing result in the limit
of zero relative inclination. The classical planetary perturbation
theory, while valid to all orders in $\alpha$, applies only to orbits
of low-mass objects orbiting a common central mass, with low
eccentricities and low relative inclination.  For triple systems
containing a close inner binary, we also discuss the possible
interaction between the classical Newtonian perturbations and the
general relativistic precession of the inner orbit. In some cases we
show that this interaction can lead to resonances and a significant
increase in the maximum amplitude of eccentricity perturbations.  We
establish the validity of our analytic expressions by providing
detailed comparisons with the results of direct numerical integrations
of the three-body problem obtained for a large number of
representative cases. In addition, we show that our expressions reduce
correctly to previously published analytic results obtained in various
limiting regimes.  We also discuss applications of the theory in the
context of several observed triple systems of current interest,
including the millisecond pulsar PSR B1620$-$26 in M4, the giant
planet in 16 Cygni, and the protostellar binary TMR-1.

\end{abstract}


\section{Introduction}

About one third of all binary star systems are thought to be members
of larger multiple systems. Most of these are hierarchical triples, in
which the (inner) binary is orbited by a third body in a much wider
orbit (see Tokovinin 1997a,b for recent results and compilations).
Secular perturbations in triples result from the gravitational
interaction between the inner binary and the outer object, possibly
coupled to other processes such as stellar evolution, tidal effects,
or, for compact objects, general relativistic effects. In strongly hierarchical
triples, the two orbits never approach each other closely, and an
analytic, perturbative approach can be used to calculate the evolution
of the system.  One particularly important perturbation is that of the
orbital eccentricities.  As the two orbits torque each other and
exchange angular momentum, their eccentricities will undergo periodic
oscillations over secular timescales (i.e., very long compared to the
orbital periods). For non-coplanar systems, corresponding oscillations
occur in the orbital inclinations.
In contrast, according to canonical perturbation theory, there is no
secular change in the semimajor axes, since the
energy exchange between the two orbits averages out to zero over
long timescales (see, e.g., Heggie 1975).

For triple systems that begin their life near the stability limit,
the result of an eccentricity increase can be catastrophic, leading to
a collision between the two inner stars, if they started as a close pair,
 or, more typically, to the disintegration of the triple. This
disintegration proceeds through a phase of chaotic evolution whose
outcome is the ejection of one of the three stars (typically the
least massive body) on an unbound trajectory, while the other two are left in a
more tightly bound binary.  A striking example of this process was
revealed by the recent HST/NICMOS observations of the TMR-1 system in
the Taurus star-forming region (Terebey \etal 1998). The HST images
reveal a faint companion, most likely a giant planet or brown dwarf,
that appears to have been ejected from its parent protostellar binary
system.  More indirect observational evidence is provided in the form
of binary systems with anomalously high space velocities. In
particular, the disintegration of short-lived triples formed in dense
star forming regions may lead to binary OB runaway stars with very
large peculiar velocities, such as HD 3950 (Gies \& Bolton 1986).

The stability of triple systems has been the subject of many
theoretical studies. Most recently, Eggleton \& Kiseleva (1995
and references therein)
performed numerical experiments and provided an empirical stability
criterion in terms of a critical ratio $Y_{\rm min}$ between the
periastron distance of the outer orbit to the apastron distance of
the inner orbit.  For systems containing three nearly equal masses,
one finds $Y_{\rm min}\simeq3-6$ depending on initial phases, eccentricities
and inclinations.  
Holman \& Wiegert (1999) study the stability of planets in
binary systems, both for planets orbiting close to one of the two
stars, and for planets orbiting outside the binary. All these
stability analyses are based on numerical integrations of the
three-body problem that are limited to $10^4-10^6$ periods of the outer
orbit.  In some cases, however, the secular evolution timescale of the
triple can be much longer than this, and therefore systems that remain
stable for the duration of the numerical integration may in fact turn
out to be unstable over secular timescales.  Analytic results such as
those derived here can therefore help determining more accurate
stability criteria, e.g., by integrating the secular evolution
equations and verifying that the stability ratio remains $>Y_{\rm
min}$ over the entire cycle of secular perturbations.

Hierarchical triple star systems can play an important role in the
dynamical evolution of dense star clusters containing primordial
binaries. The cores of globular clusters, for example, are thought to
contain a small but dynamically significant population of triple
systems formed through dynamical interactions between primordial
binaries (McMillan, Hut, \& Makino 1991). Both stable and unstable
triples can form easily through exchange and resonant interactions
between binaries. In direct $N$-body integrations of the cluster
dynamics, marginally stable or unstable triples can represent a
significant computational bottleneck, since they require very long
integrations of the orbital dynamics in order to resolve the outcome
of the interaction (see, e.g., Mikkola 1997).

Direct observational evidence for the dynamical production of triple
systems in globular clusters is provided by the millisecond pulsar
system PSR B1620$-$26 (Rasio, McMillan, \& Hut 1995; Ford \etal 2000). 
This radio pulsar is a member of
 a hierarchical triple system located in the core of the globular
cluster M4. The inner binary of the triple contains the $\simeq 1.4
M_{\odot}$ neutron star with a $\simeq 0.3 M_{\odot}$ white-dwarf
companion in a 191-day orbit (Lyne \etal 1988; McKenna \& Lyne 1988).
The triple nature of the system was first proposed by Backer (1993) in
order to explain the unusually high residual second and third pulse
frequency derivatives left over after subtracting a standard Keplerian
model for the pulsar binary.  The pulsar has now been timed for eleven
years since its discovery (see Thorsett \etal 1999 for the most recent
update).  These observations have not only confirmed the triple nature
of the sytem, but they have also provided tight constraints on the
mass and orbital parameters of the second companion.  Theoretical
modeling of the latest timing data (now including five pulse frequency
derivatives) and preliminary measurements of the orbital perturbations
of the inner binary have further constrained the mass of the second
companion, and strongly suggest that it is a giant planet or a brown
dwarf of mass $\sim 0.01\,M_\odot$ at a distance of $\sim 50\,$AU
from the pulsar binary (Joshi \& Rasio 1997; Ford \etal 2000).

Our treatment of the secular perturbations in this paper is based on
classical celestial mechanics techniques and assumes that all three
bodies are unevolving point masses. Whenever the stellar evolution
time of one of the components becomes comparable to any of the orbital
perturbation timescales computed here, the evolution of the triple can
be affected significantly through mass losses, or mass transfer. For a
recent discussion of stellar evolution in triples, see Iben \& Tutukov
(1999), who study the production of Type~Ia supernovae from the mergers
of heavy white dwarfs inside hierarchical triples. 
Mikkola \& Tanikawa (1998) have studied the episodic mass transfer triggered
by large eccentricity oscillations of the inner binary in the secular
evolution of the triple system CH Cygni.
Eggleton \& Verbunt (1989) have discussed the possible
relevance of triple star evolution for the formation of low-mass X-ray
binaries.  

Our work focuses on triple systems containing
well-separated components, in which the orbital perturbation
timescales are short compared to any tidal dissipation time. For
triple systems containing a close inner binary, tidal dissipation in
the inner components provides a sink of energy and angular momentum
which can change substantially the character of the secular
perturbations.  For a recent discussion of tidal dissipation in triple
systems containing a close inner binary, see the paper by Kiseleva,
Eggleton, \& Mikkola (1998).  Bailyn \& Grindlay (1987) have discussed
the combined effects of tidal interaction and mass transfer for
compact X-ray binaries in hierarchical triples.
Mazeh \& Shaham (1979) were the first to point out that the combination of
tidal dissipation and secular eccentricity perturbations in triples
could sometimes lead to a substantial orbital shrinking of the
inner binary.

One possible additional perturbation effect that we do take into
account in this work is the general relativistic precession of the
inner orbit if the inner binary contains compact objects. An example
is provided by the PSR B1620$-$26 triple system, in which the inner
binary contains a neutron star and a white dwarf. When the precession
periods from general relativity and from Newtonian perturbations in
the triple become comparable, a type of resonant effect is possible
which leads to increased magnitudes for the orbital perturbations.  A
similar resonant effect has been mentioned by S\"oderhjelm (1984) for
triples where the inner binary precesses under the influence of a
rotationally-induced quadrupole moment in one of the stars.

Our paper is organized as follows.  In \S 2 we present a derivation of
the octupole-order secular perturbation equations and we
compare our results to those obtained in the quadrupole approximation
and in classical planetary perturbation theory.  In \S 3 
the analytic results are compared to direct numerical integrations, and
the effects of varying all relevant parameters are explored.
In \S 4 we discuss the effects of the general relativistic precession of the
inner orbit on the secular evolution of the triple, using 
the PSR B1620$-$26 system as an example.
In \S 5 possible applications of our results to
other observed triple systems are briefly discussed.

\section{Analytic Secular Perturbation Theory}

In this section we present a simple analytic treatment of the
long-term, secular evolution of hierarchical triple systems using
time-independent Hamiltonian perturbation theory in which the
small parameter is the ratio of semimajor axes. We discuss essential
aspects of the lowest-order (quadrupole-level) approximation, which
has been widely used to study hierarchical stellar triples. Then we
extend the approximation to the octupole level and compare our results
with results from the quadrupole and other approximations to show that
the octupole-level equations derived here are valid for a far greater
range of parameters.

\subsection{Summary of Previous Work}

Our derivation of the octupole-level secular perturbation equations is
based on classical perturbation methods of celestial mechanics.
Studies of the long-term behavior of the solar system led Lagrange and
Laplace to the creation of the first classical perturbation theory.
Their approach is applicable to a small class of planetary
configurations with parameters similar to those of the solar system.
The lunar problem was successfully attacked in the end of the
last century by Delaunay who was the first to apply the method of
canonical transformations to long-term perturbations. This method
possesses much greater generality and was used to study a broad
spectrum of problems.  Brown (1936) was the first to apply canonical
averaging to stellar triples, and he obtained the transformed
quadrupole Hamiltonian.  Kozai (1962) made use of the quadrupole
approximation while studying the long-term motion of asteroids and
noted several important properties of this approximation. Harrington
(1968) obtained quadrupole-level expressions similar to Kozai's for
general hierarchical systems of three stars.  S\"oderhjelm (1984)
derived octupole-level equations in the limit of low eccentricities
and inclinations. In particular, he demonstrated that the quadrupole
approximation fails in this regime because the octupole term in the
Hamiltonian becomes dominant. Finally, Marchal (1990) averaged the
octupole Hamiltonian keeping all terms up to third order in
$\alpha$ and some terms of order $\alpha^{7/2}$.  His Hamiltonian
truncated at third order is identical to the one used in this paper.

In the process of completing this work, we became aware of related
ongoing work by other groups. In particular, Krymolowski \& Mazeh
(1999) have derived octupole-order perturbation equations following
the same method used here. They retain some additional 
terms, of order $\alpha^{7/2}$, which were also partly included
in Marchal's (1990) Hamiltonian.  Based on a few numerical integrations 
that Krymolowski \& Mazeh (1999) provide for
a fairly strongly coupled system ($\alpha=0.1$), it appears that
these higher-order terms have a negligible effect on 
the perturbations, although they can lead to slightly shorter
periods of eccentricity oscillations for systems with low relative
inclination. Eggleton (2000) has used a perturbation method based
on the variation of the Runge-Lenz vector (see Heggie \& Rasio 1996)
to derive an extension of Kozai's theory to octupole order.
Similar work has been done by Georgakarakos (2000), who
concentrates on systems where the inner orbit is nearly circular.

\subsection{Octupole Theory}

A hierarchical triple system
consists of a close binary ($m_0$ and $m_1$) and a third
body ($m_2$) moving around the inner binary on a much wider orbit.
To describe this structure it is convenient to
use Jacobi coordinates, which are defined as follows. The vector ${\bf r}_1$
represents the position of $m_1$ relative to $m_0$, and ${\bf r}_2$ is
the position of $m_2$ relative to the center of mass of the inner binary 
(See Fig.\ 1). 
This coordinate system naturally divides the motion
of the triple into two separate motions, and makes it possible to
write the Hamiltonian as a sum of two terms representing the two
decoupled motions and an infinite series representing the coupling
of the orbits. Let the subscripts 1 and 2 refer to the inner
and outer orbits, respectively. The coupling term is written as a
power series in the ratio of the semi-major axes
$\alpha\equiv{a_1}/{a_2}$, which serves as the small parameter
in our perturbation expansion. The complete Hamiltonian of the
three-body system is given by (Harrington 1968)
\beq
    \Hm=\frac{k^2 m_0 m_1}{2 a_1} + \frac{k^2 (m_0+m_1) m_2}{2a_2}+
    \frac{k^2}{a_2}\, \sum_{j=2}^\infty \alpha^{j} M_j \left(
    \frac{r_1}{a_1} \right)^j \left(\frac{a_2}{r_2}\right)^{j+1}
P_j(\cos \Phi) \label{eq:Ham1},
\feq
where $k^2$ is the gravitational constant, $P_j$'s are the Legendre
polynomials, $\Phi$ is the angle between ${\bf r}_1$ and ${\bf r}_2$, and
\beq 
M_j = m_0m_1m_2\frac{m_0^{j-1}-(-m_1)^{j-1}}{(m_0+m_1)^j} .
\feq
We shall deal with the expansion only up to third order in $\alpha$.

\begin{figure}
\plotone{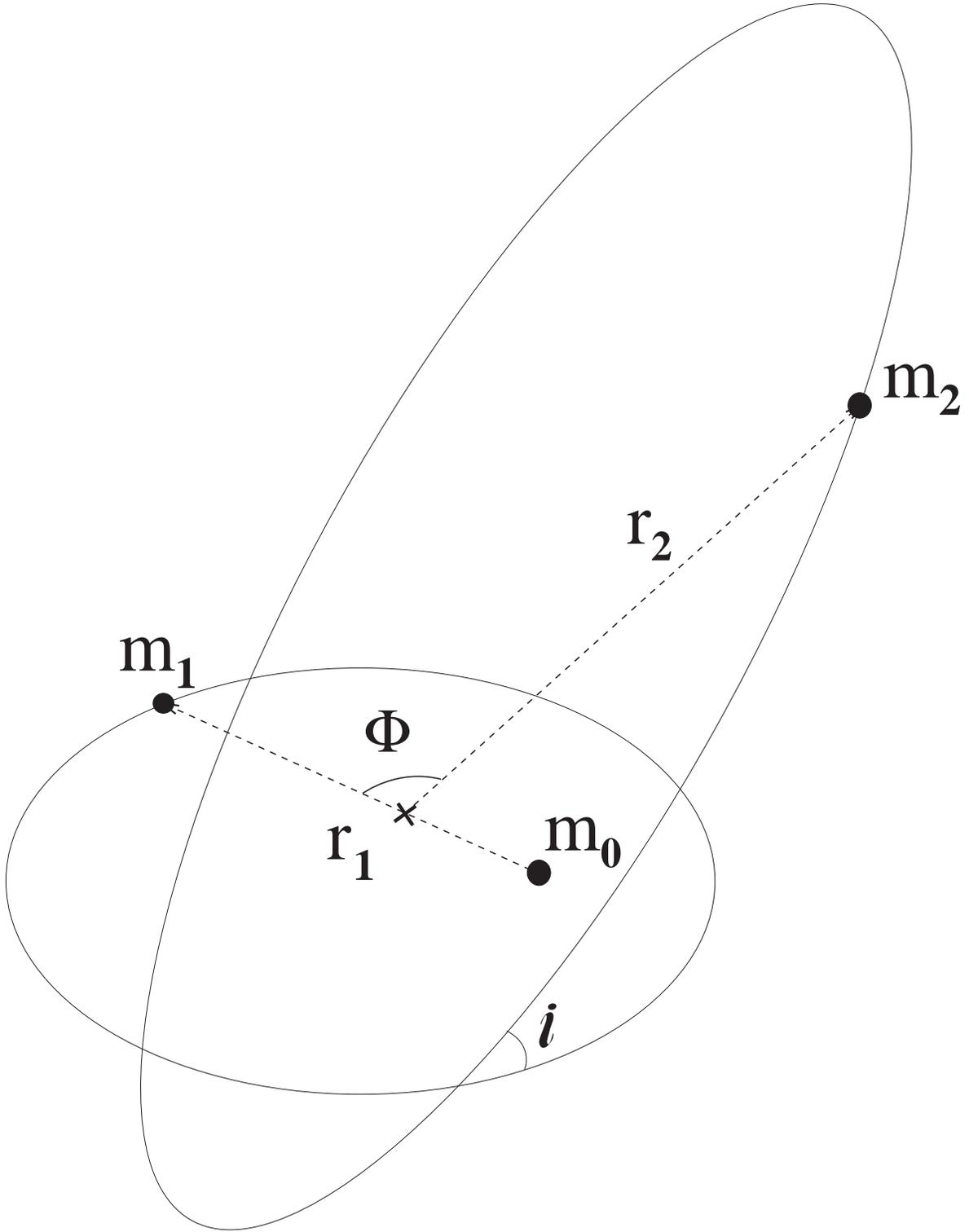}
\caption{Diagram illustrating the coordinate system used to describe the 
hierarchical triple system. \label{fig:coord}}
\end{figure}

Let us define a set of canonical
variables, known as Delaunay's elements, that provide a particularly
convenient dynamical description of our three-body system. The angle
variables are chosen to be
\beq 
l_1,l_2 &=& \rm {mean \ anomalies} \\ g_1,g_2 &=& \rm {arguments \
   of \ periastron} \\ h_1,h_2 &=& \rm{longitudes \ of \ ascending \
nodes} \label{eq:def1} 
\feq
and their conjugate momenta
\beq 
L_1 = \frac{m_0m_1}{m_0+m_1}\sqrt{k^2(m_0+m_1)a_1} \qquad
 L_2=\frac{m_2(m_0+m_1)}{m_0+m_1+m_2}\sqrt{k^2(m_0+m_1+m_2)a_2},
\feq
\beq 
G_1 = L_1 \sqrt{1-e_1^2} \qquad G_2=L_2 \sqrt{1-e_2^2},
\feq
\beq 
H_1 = G_1 \cos i_1 \qquad H_2=G_2 \cos i_2, \label{eq:def2}
\feq
where $e_1$, $e_2$ are the orbital eccentricities and $i_1$, $i_2$ are the
orbital inclinations.

The usual canonical relations represent the equations of motion:
\beq  
\frac{dL_j}{dt}=\frac{\partial \Hm}{\partial l_j} \qquad
 \frac{dl_j}{dt}=-\frac{\partial \Hm}{\partial L_j}, \label{eq:canon1a} 
\feq
\beq  
\frac{dG_j}{dt}=\frac{\partial \Hm}{\partial g_j} \qquad
 \frac{dg_j}{dt}=-\frac{\partial \Hm}{\partial G_j} ,
\feq
\beq  
\frac{dH_j}{dt}=\frac{\partial \Hm}{\partial h_j} \qquad
 \frac{dh_j}{dt}=-\frac{\partial \Hm}{\partial H_j}, \label{eq:canon1b} 
\feq
where $j=1,2$.
Note that $H_2$ is the $z$-component of the angular momentum contributed by
the perturbing body, the $z$-axis being the direction of the total angular
momentum, perpendicular to the invariable plane of the system (See Fig.\ 2).
Equations~(\ref{eq:canon1a})--(\ref{eq:canon1b}) appear to
have six degrees of freedom, but they can be reduced to four by the theorem 
of elimination of nodes (Jeffrys \& Moser 1966). The Hamiltonian contains
$h_1$ and
$h_2$ only in the combination $h_1-h_2$, and it is symmetric with respect to 
the orientation of the line of nodes when the invariable plane is chosen 
as a reference plane. In other words, the Hamiltonian is symmetric with
respect to rotations about the total angular momentum vector $\bf H$.
Thus, $H_1$ and $H_2$ enter the Hamiltonian only as $H_1+H_2=H$ and can
be eliminated from the Hamiltonian using the relations
\beq
H_1=\frac{H^2+G_1^2-G_2^2}{2H} \\
H_2=\frac{H^2+G_2^2-G_1^2}{2H}
\feq

\begin{figure}
\plotone{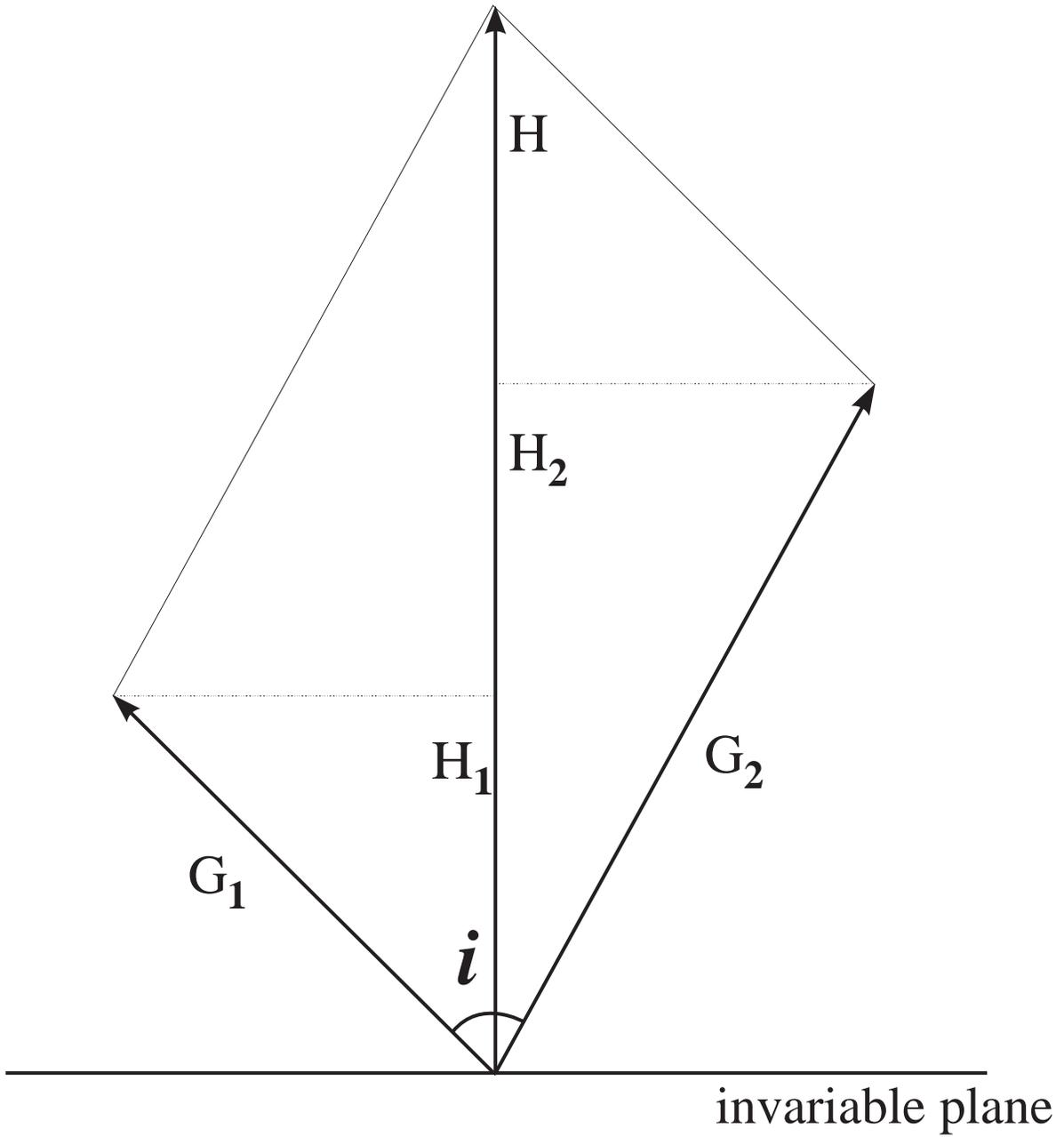}
\caption{Diagram illustrating the relationships between the canonical
variables and angular momenta. \label{fig:vectors} }
\end{figure}

Using the new canonical elements we can write the first four terms of 
the Hamiltonian~(\ref{eq:Ham1}) as
\beq 
  \Hm_{oct} &=& \Hm_0 +\Hm_1 +\Hm_2 +\Hm_3 \\ \label{eq:Ham2}
  &=&\frac{\beta_0}{2L_1^2}+\frac{\beta_1}{2L_2^2}+8\beta_2 
  \left(\frac{L_1^4}{L_2^6}\right) \left(\frac{r_1}{a_1}\right)^2
  \left(\frac{a_2}{r_2}\right)^3(3\cos^2\Phi-1) \\ 
  & & + 2 \beta_3
  \left(\frac{L_1^6}{L_2^8}\right) \left(\frac{r_1}{a_1}\right)^3
  \left(\frac{a_2}{r_2}\right)^4(5\cos^3\Phi-3\cos\Phi), \nonumber
\feq
where the mass parameters are
\beq
  \beta_0 &=& k^4 \frac{(m_0m_1)^3}{m_0+m_1}, \\
  \beta_1 &=& k^4 \frac{(m_0+m_1)^3m_2^3}{m_0+m_1+m_2}, \\
  \beta_2 &=& \frac{k^4}{16}\frac{(m_0+m_1)^7}{(m_0+m_1+m_2)^3}
              \frac{m_2^7} {(m_0m_1)^3},\\
  \beta_3 &=& \frac{k^4}{4}\frac{(m_0+m_1)^9}{(m_0+m_1+m_2)^4}
              \frac{m_2^9(m_0-m_1)}{(m_0m_1)^5}.
\feq
Each term in the series is labeled according to the degree of the Legendre
polynomial associated with it (same as the power of $r_1/r_2$). Following the
standard 
nomenclature associated with multipole expansions we shall call the 
Hamiltonian containing the first three terms (up to $j=2$) ``quadrupole'' and
the third-order Hamiltonian~(\ref{eq:Ham2}) ``octupole.'' 
The first two terms in expression~(\ref{eq:Ham2}) describe the unperturbed motion
of the inner and outer binaries, and the higher-order terms describe the coupling.
The quadrupole
Hamiltonian contains the perturbation of order $\alpha^2$, the octupole Hamiltonian
extends this to order $\alpha^3$.   

The complete Hamiltonian~(\ref{eq:Ham1})  contains the full description
of the system. However, we are going to restrict our study to the
long-term, secular behavior of the system, by averaging over short-period
effects. 
Even though equation~(\ref{eq:Ham2}) is already an approximation of the
full Hamiltonian, it contains information about short-period perturbations
that needs to be eliminated. In particular the angle 
$\Phi$ depends on the mean anomalies. Further simplification is achieved 
through a canonical transformation of variables,
called the von Zeipel transformation. 
Its essence is to replace the Delaunay elements with a set of new canonical
coordinates and momenta that rid the Hamiltonian of the dependence
on $l_1$ and $l_2$. The perturbed action variables are still periodic
functions of the perturbed angle variables, but the former are no
longer linear functions of time. The goal is to find such a set of
action-angle variables that the perturbed Hamiltonian will be a
function only of the action variables.
In the end we can think of the Hamiltonian as describing the
interaction between two weighted elliptical rings instead of point masses
in orbits. 

It is important to note that we do not
simply average the Hamiltonian with respect to these variables,
since this would destroy the canonical structure of the equations 
of motion, but instead proceed in a more cautious and intricate manner.
We start by requiring that the new Hamiltonian be equal to the old one,
since changing variables does not change the energy, and
expanding both sides of the equality as Taylor series in $\alpha$.
Then we go order by order to identify the terms in the transformed
Hamiltonian, using the result of the previous order calculation in each step.
The theory behind the Von Zeipel method is very well presented and
illustrated by Goldstein (1980, section 11-5) and Hagihara (1972).  
Additionally, Harrington (1968, 1969)
has applied this method to the quadrupole Hamiltonian.  We
followed exactly the same prescription but all the way to third order.
Here we present only the results of the von Zeipel averaging
procedure, omitting the laborious algebraic details. 

Let us define the following convenient quantities
\beq 
\theta=\cos i = \frac{H^2-G_1^2-G_2^2}{2 G_1 G_2}, \label{eq:cosi} 
\feq
where $\bf H= \bf G_1 + \bf G_2$ and $H=|\bf H|$ is given by
initial conditions and $i=i_1-i_2$ is the mutual inclination.
The angle $\varphi$ between the directions of periastron is given by
\beq 
\cos \varphi= -\cos g_1 \cos g_2-\theta \sin g_1 \sin g_2.
\feq
The doubly-averaged Hamiltonian is given by
\beq
     \bar \Hm_{oct} &=& C_2 \cdot \{(2+3e_1^2)(3\theta^2-1)+
     15e_1^2(1-\theta^2)\cos 2g_1 \} \nonumber\\
     & & +C_3 \cdot e_1 e_2 \Big\{A \cos \varphi + 10\theta(1-\theta^2)(1-e_1^2)
     \sin g_1 \sin g_2 \Big\}, \label{eq:Ham3}
\feq
where
\beq 
 C_2=\frac{k^4}{16}\frac{(m_0+m_1)^7}{(m_0+m_1+m_2)^3}
       \frac{m_2^7}{(m_0m_1)^3}
       \frac{L_1^4}{L_2^3G_2^3}, \label{eq:c2}
\feq
\beq 
 C_3= \frac{15}{16}\frac{k^4}{4}\frac{(m_0+m_1)^9}{(m_0+m_1+m_2)^4}
        \frac{m_2^9(m_0-m_1)}{(m_0m_1)^5}
        \frac{L_1^6}{L_2^3G_2^5},  \label{eq:c3}
\feq
\beq B=2+5e_1^2-7e_1^2\cos 2g_1,\feq
\beq A=4+3e_1^2-\frac{5}{2}(1-\theta^2)B.\feq
Note that the Hamiltonian~(\ref{eq:Ham3}) does not contain any
dependence on $l_1$ or $l_2$ because these variables have been
integrated out as a result of the canonical transformation.
The actual differences between the
original and the transformed variables are small (of order $\alpha$ or
smaller) and periodic. Variables
appearing in equation~(\ref{eq:Ham3}) are only approximations of the old
variables defined in equations~(\ref{eq:def1})--(\ref{eq:def2}), 
and they can be thought of
as the averages of the old variables.  This Hamiltonian is equivalent to 
the Hamiltonian given by Marchal (1990; see his eqs.~[252]--[255]).

The absence of $l_1$ and $l_2$ from the transformed Hamiltonian implies
that $L_1$ and $L_2$ are constants of the motion, which in turns implies that,
 in our approximation, the transformed semi-major axes $a_1$ and $a_2$ are
constant. Thus, the only
secularly changing parameters in this model are $e_1$, $e_2$, $g_1$, 
$g_2$, and $i$, which is coupled to $e_1$ and $e_2$ by
relation~(\ref{eq:cosi}).
The equations of motion are derived from the Hamiltonian~(\ref{eq:Ham3}) 
using the canonical relations 
\beq \frac{de_i}{dt}=\frac{\partial e_i}{\partial G_i}
   \frac{\partial  \bar \Hm_{oct}}{\partial g_i}, \label{eq:canon2}
\feq
and
\beq 
\frac{dg_i}{dt}=-\frac{\partial \bar \Hm_{oct}}{\partial G_i}.
\feq

After regrouping terms, the octupole-level secular perturbation equations 
follow:
\beq
    \frac{dg_1}{dt} &=& C_2 \cdot 6 \left\{\frac{1}{G_1}\left[4 \theta^2+(5
\cos 2g_1
    -1)(1-e_1^2-\theta^2)\right]+ \frac{\theta}{G_2} \left[2+e_1^2(3-5
\cos2g_1)\right]
    \right\} \nonumber \\ & & -C_3\cdot e_2 \Bigg\{e_1\left(
    \frac{1}{G_2}+ \frac{\theta}{G_1}\right) \Big[\sin g_1 \sin g_2 \left\{
    A+10(3\theta^2-1)(1-e_1^2)\right\}
    -5\theta B \cos \varphi \Big] \nonumber \\& &
    - \frac{1-e_1^2}{e_1G_1}\Big[\sin g_1 \sin g_2 \cdot 10 \theta
    (1-\theta^2)(1-3e_1^2) + \cos \varphi (3A-10\theta^2+2)
    \Big]\Bigg\}, \label{eq:oct1a} 
\feq
\beq
    \frac{de_1}{dt} &=& C_2 \cdot \frac{1-e_1^2}{G_1} \Big\{
    30e_1 (1-\theta^2)\sin 2g_1\Big\} + C_3 \cdot e_2\frac{1-e_1^2}{G_1}
    \Big\{35 \cos \varphi (1-\theta^2)e_1^2\sin 2g_1 \nonumber \\
    & & - 10\theta(1-e_1^2)(1-\theta^2) \cos g_1 \sin g_2 -
    A(\sin g_1 \cos g_2-\theta \cos g_1 \sin g_2)\Big\},
\feq
\beq
    \frac{dg_2}{dt} &=& C_2 \cdot 3\left\{
\frac{2\theta}{G_1}\left[2+e_1^2(3-5\cos 2g_1)
    \right] + \frac{1}{G_2} \left[4+6e_1^2+
    (5\theta^2-3)(2+e_1^2(3-5\cos 2g_1))\right]\right\} \nonumber \\
    & & + C_3\cdot e_1 \Bigg\{\sin g_1 \sin g_2 \left[\frac{4e_2^2+1}{e_2G_2}
    10 \theta (1-\theta^2)(1-e_1^2)- e_2 \left(\frac{1}{G_1}+
    \frac{\theta}{G_2}\right)(A+10(3\theta^2-1)(1-e_1^2))\right] \nonumber \\
    & & +\cos \varphi \left[ 5B\theta e_2\left(\frac{1}{G_1}+
    \frac{\theta}{G_2}\right)+\frac{4e_2^2+1}
    {e_2G_2}A \right]\Bigg\},
\feq
\beq
    \frac{de_2}{dt} &=& -C_3 \cdot e_1 \frac{1-e_2^2}{G_2} \Big\{
    10\theta(1-\theta^2)(1-e_1^2)\sin g_1 \cos g_2 + A(\cos g_1 \sin g_2-
     \theta \sin g_1 \cos g_2) \Big\}. \label{eq:oct1b} 
\feq
This system of coupled nonlinear differential equations describes
the octupole-level behavior of hierarchical triples and can be easily
integrated numerically to determine the secular evolution of a
hierarchical triple for any initial configuration.  We found that for
calculations with small eccentricities it is better to use the
transformed set of variables $e_1 \sin g_1$, $e_1 \cos g_1$, $e_2 \sin
g_2$, and $e_2 \cos g_2$.  These equations can be integrated
numerically much more rapidly than the full equations of motion.

\subsection{Comparison with Quadrupole-Level Results}

The quadrupole results of Kozai (1962) and Harrington (1969) have
found numerous applications in recent studies of planetary, pulsar,
and stellar systems. Mazeh \& Shaham (1979) calculated the
quadrupole-level, long-term periodic behavior of triples and used
their results to study high-amplitude eccentricity modulations of
systems with large initial inclination.  Holman et al.\ (1997) used
the quadrupole-level theory of Kozai to analyze the dynamical
evolution of the planet in the binary star 16 Cygni.  Rasio, Ford, \&
Kozinsky (1997) used the same theory to study long-term eccentricity
perturbations in the PSR B1620$-$26 pulsar system.

The quadrupole Hamiltonian can be obtained from expression~(\ref{eq:Ham3}) by
dropping the $C_3$ term:
\beq
\Hm_q=C_2 \cdot \{(2+3e_1^2)(3\theta^2-1)+ 
  15e_1^2(1-\theta^2)\cos 2g_1 \}.
\feq 
Note that $\Hm_q$ is independent of $g_2$, meaning that, in the quadrupole
approximation, $G_2$ is a constant of the motion, and so is $e_2$ by
equation~(\ref{eq:canon2}). Therefore, to quadrupole order in secular
perturbation theory, there is no variation in the eccentricity of the 
outer orbit. This is a well-known result (see, e.g., Marchal 1990,
Sec.~10.2.3). Now there is only one degree of freedom left, so the
evolution of the inner eccentricity is described by
\beq
    \frac{dg_1}{dt} &=& C_2 \cdot 6 \left\{\frac{1}{G_1}\left[4 \theta^2+(5
\cos 2g_1
    -1)(1-e_1^2-\theta^2)\right]+ \frac{\theta}{G_2} \left[2+e_1^2(3-5
\cos2g_1)\right]
    \right\}, \label{eq:quad1a} 
\feq
with
\beq
    \frac{de_1}{dt} &=& C_2 \cdot \frac{1-e_1^2}{G_1} \Big\{
    30e_1 (1-\theta^2)\sin 2g_1\Big\}. \label{eq:quad1b} 
\feq

The advantage of using the quadrupole approximation is that it can
describe the secular behavior of systems with high relative
inclination and a wide range of initial eccentricities, regimes not
covered by the classical planetary perturbation theory.  As in the
classical theory, the quadrupole-level perturbation equations
(\ref{eq:quad1a})--(\ref{eq:quad1b}) can be solved exactly for the 
period and amplitude of oscillation (Kozai 1962). The period of
eccentricity oscillations is given approximately by
\beq
P_e \simeq P_{1} \left(\frac{m_0+m_1}{m_2}\right)
\left(\frac{a_2}{a_1}\right)^3 \left( 1- e_2^2 \right)^{3/2},
\feq
where $P_1$ is the orbital period of the inner binary
(Mazeh \& Shaham 1978).
This expression should be multiplied by a coefficient of order unity 
which can be obtained using
Weierstrass's zeta function as shown by Kozai (1962).

The secular evolution can be visualized with the help of phase-space
diagrams of $e_1$ vs $\cos g_1$. An example is provided in Figure~3.
Each contour corresponds to an initial condition with a certain value
of the total angular momentum $H$. Since $G_2$ is fixed, $e_1$ is
coupled to $i$ through equation~(\ref{eq:cosi}), so the relative
inclination oscillates with the same period as $e_1$.  The up-down
symmetry forces similar behavior in $g_1$-$e_1$ phase space for both
$g_1 \in [-\pi,0]$ and $g_1 \in [0,\pi]$.  One obvious feature is the
existence of two regimes: libration and circulation.  The libration
island generally appears when the initial inclination is greater than
some critical value, which for most systems is around
$i_{\rm crit}\simeq 40^{\circ}$. 
Kozai (1962) calculated that for $\alpha \le 0.10$,
$\,38.960^{\circ} \le i_{\rm crit} \le 39.231^{\circ}$.  From the shape
of the large libration island we see that $e_1$ can grow from a very
small initial value to a very large maximum. Holman \etal (1997)
approximate the maximum inner eccentricity as
\beq
e_{1, {\rm max}} \simeq \left(1-(5/3) \cos^2 i_o\right)^{1/2},
\feq
where $i_o$ is the initial relative inclination. 
Libration can occur for low inclinations as well, but this does 
not lead to large eccentricity oscillations. 

Several erroneous features of the quadrupole approximation are worth
noting.  For an initial condition with $e_1=0$, no evolution occurs at
all. This is especially significant in a case where there is no
libration island, since in that case the eccentricity perturbation
would appear to
approach zero continuously as the initial eccentricity is decreased.
Similarly, in the coplanar case ($\theta=1$), the theory predicts no
evolution of eccentricity. From the classical planetary
perturbation theory (\S 2.4), which assumes low eccentricities and
inclinations, we know that this is not correct. These features also
contradict the octupole-level results (\S 2.2), as well as the results
of direct numerical integrations (\S 3.2). We conclude that the
quadrupole approximation fails for low inclination and for low inner
eccentricity.  However, it remains qualitatively applicable in the
high-inclination regime.

\begin{figure}
\vspace{6.5in}
\includegraphics{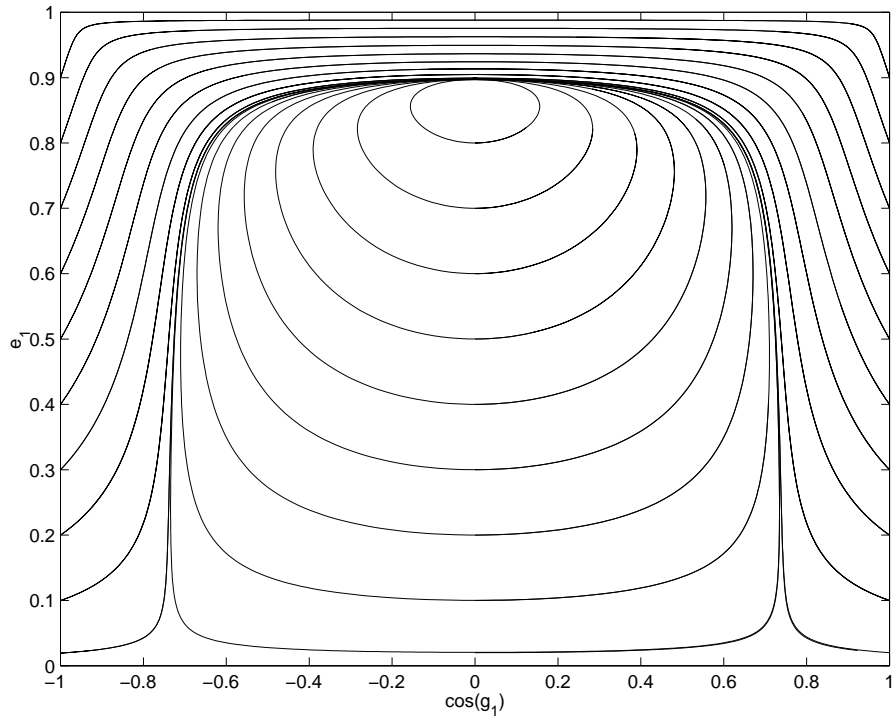}
\vspace{-1in}
\caption{Phase space trajectories obtained in the quadrupole
approximation for a system with  $m_2/m_1 = 10^{-3}$, $m_3/m_1 =
1$, $\alpha^{-1} = 100$, $e_2 = 0.9$, and initial values of $e_1$ ranging
from $0.02$ to $0.9$. The libration contours were obtained by setting
the initial value of $g_1=90^{\circ}$.\label{fig:phase1}}
\end{figure}

\begin{figure}
\vspace{6.5in}
\includegraphics{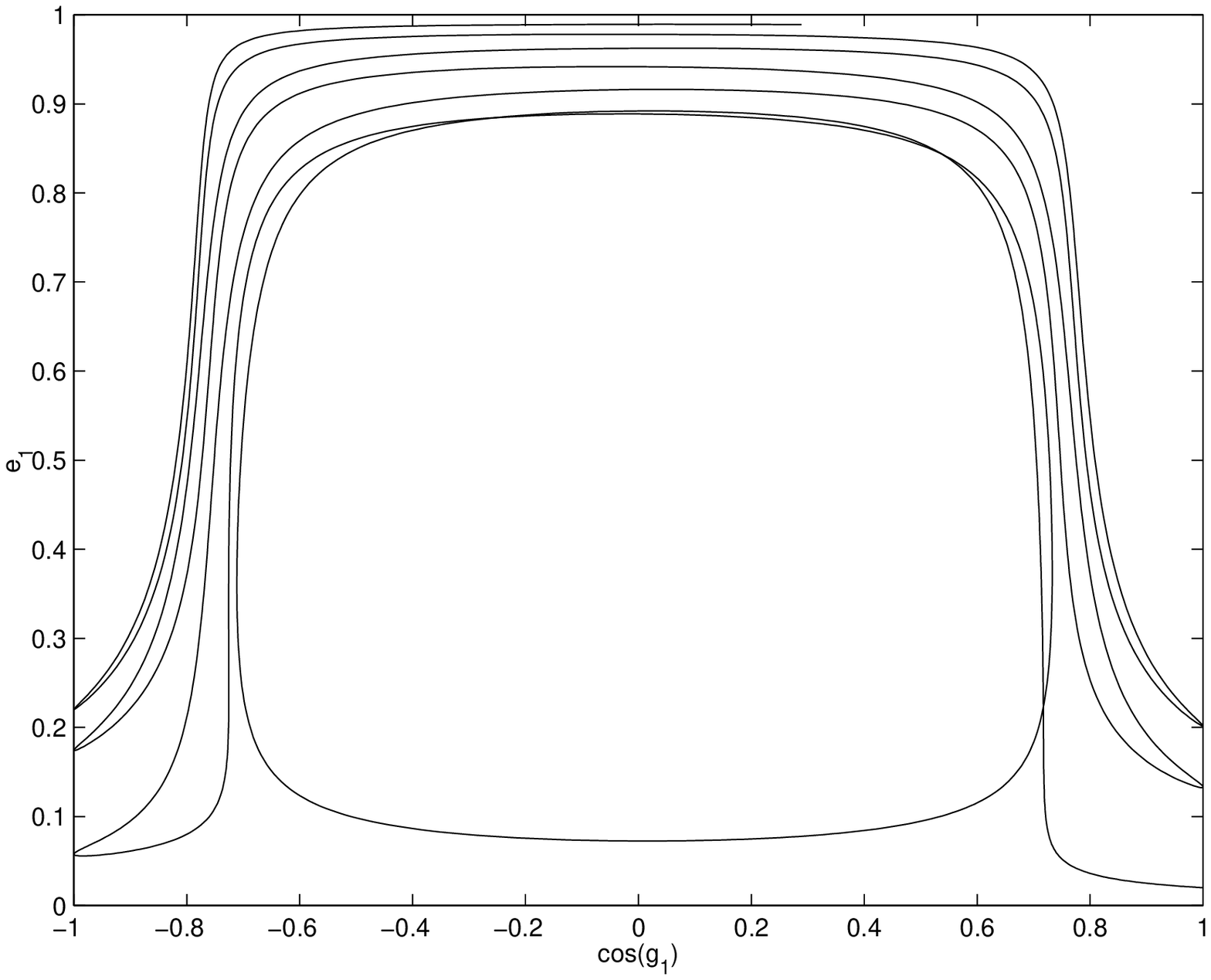}
\vspace{-1in}
\caption{Phase-space trajectory of the same system as in Figure~3 with initial 
$e_1=0.02$, but with the octupole terms included in the integration. 
(Note that this corresponds to a single trajectory in Fig.~3.)
\label{fig:phase2}}
\end{figure}

The octupole-level theory has more degrees of freedom and covers most
regimes of hierarchical triple configurations. The perturbation
equations~(\ref{eq:oct1a})--(\ref{eq:oct1b}) indicate that there are no
additional conserved quantities apart from the obvious ones (total
angular momentum and total energy). In contrast to the quadrupole theory, the
quantities $e_2$ and $g_2$ now vary with time and the behavior is much
harder to visualize. We can notice striking qualitative
differences between the two theories by looking at phase-space
diagrams. An example is provided in Figure~\ref{fig:phase2}. Trajectories
are no longer closed, and transitions between libration and
circulation occur, since the angular momentum of the outer orbit now
evolves with time.  Thus, we now have more
than one frequency in the secular oscillations. 

\begin{figure}
\plotone{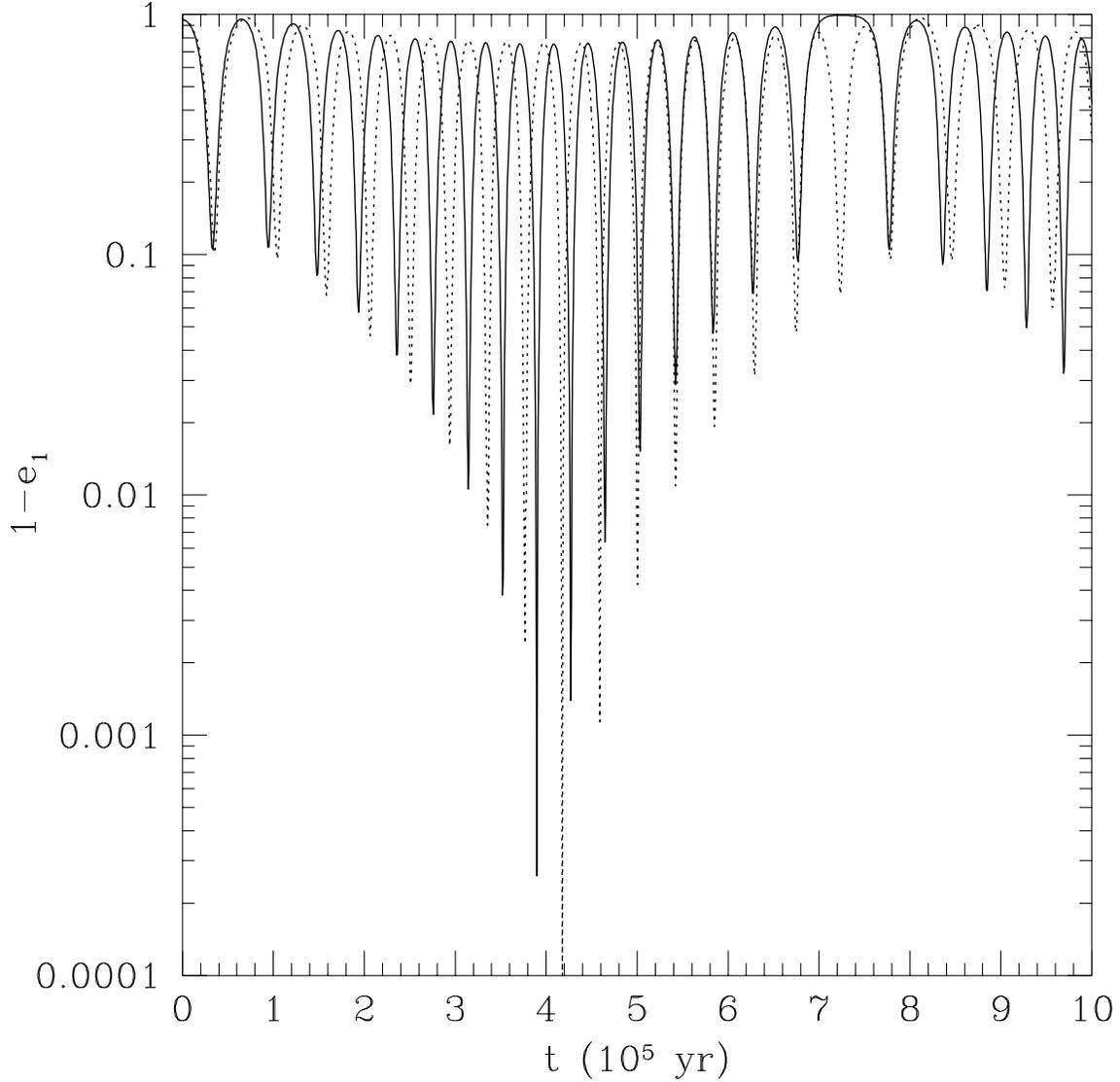}
\caption{These results illustrate the potential danger of
using the quadrupole approximation.
The inner eccentricity is shown as a function of time for 
a system with $m_1/m_0 = 10^{-3}$, $m_2/m_1 = 1$,
$\alpha^{-1} = 100$, initial eccentricities $e_1 = 0.05$
and $e_2 = 0.9$, and an initial relative inclination $i=70^{\circ}$.  
Time is given in years assuming $a_1=1\,$AU and $m_0=1\,M_\odot$.
The solid line
is from the integration of the octupole-level perturbation equations, while
the dashed line is from a direct numerical integration of the three-body
system (see \S 3.1).  In the quadrupole approximation all oscillations would
have the same amplitude as the first shown here.  Notice how the eccentricity
oscillations in fact increase in amplitude, making the inner
periastron separation quite small.  At some point other effects such as
general relativistic precession and tidal dissipation (not to mention
a collision between the two inner stars) could become significant. 
\label{fig:bad} }
\end{figure}

Figure~\ref{fig:bad}
demonstrates that a system can have a qualitatively different behavior
from what is expected in the quadrupole approximation.  While the 
quadrupole theory predicts periodic variations of constant amplitude, 
according to the octupole equations (and in agreement with the results of
a direct numerical integration) the
amplitude grows very close to unity.  This leads to a very small periastron
distance and the possibility of a tidal interaction or collision between 
the two inner stars.
Thus, one must exercise great caution when modeling systems using 
the quadrupole approximation.  Ignoring octupole-level terms can 
lead to completely invalid results.

\subsection{Comparison with Classical Planetary Perturbation Theory}

For application to many problems in the context of the Solar System, a
classical perturbation theory was developed many years ago that
applies to low-eccentricity, low-inclination orbits of planets around
a central star (one dominant mass). This theory does {\it not\/}
assume that the ratio of semimajor axes is small, and therefore it
provides results valid to all orders in $\alpha$.  A detailed account
of the planetary theory can be found in Brouwer \& Clemence (1961,
Chap.\ 16). For an excellent pedagogic summary, see Dermott \&
Nicholson (1986). Rasio (1994, 1995) used the classical theory to
study the eccentricity perturbations in the PSR B1620-26 triple system
in the limit of coplanar orbits, and derived simple approximate
expressions for the period and amplitude of eccentricity oscillations
in various limits.

Here we will only consider the variations of the
eccentricities, since the results for inclinations are very
similar. 
It turns out that the inclination evolution is decoupled from the
eccentricity evolution, and so the two can be solved separately. 
Because eccentricities are very small and can vanish, it is
better to use the variables
\beq h_1=e_1 \sin g_1 \qquad h_2=e_2 \sin g_2  \label{eq:h} \feq
\beq k_1=e_1 \cos g_1 \qquad k_2=e_2 \cos g_2. \label{eq:k} \feq
Since angular
momentum is conserved and the mutual inclination stays
constant to first order, the two eccentricities vary $90^{\circ}$ out 
of phase. The linear
system of equations describing the secular evolution of eccentricities 
in planetary perturbation theory is
\beq \frac{dh_1}{dt}=+A_{11}k_1-A_{12}k_2 \feq
\beq \frac{dk_1}{dt}=-A_{11}h_1+A_{12}h_2 \feq
\beq \frac{dh_2}{dt}=-A_{21}k_1+A_{22}k_2 \feq
\beq \frac{dk_2}{dt}=+A_{21}h_1-A_{22}h_2, \feq
where the $A$'s are defined in terms of Laplace coefficients, which we
truncate at third order in $\alpha$ to obtain
\beq A_{11}=\frac{3}{4}k\frac{\tilde{a}_1^{3/2}}{\tilde{a}_2^3}\frac{m_2}{\sqrt{m_0+m_1}},
\qquad 
  A_{12}=\frac{15}{16}k\frac{\tilde{a}_1^{5/2}}{\tilde{a}_2^4}\frac{m_2}{\sqrt{m_0+m_1}}, \feq
\beq A_{21}=\frac{15}{16}k\frac{\tilde{a}_1^3}{\tilde{a}_2^{9/2}}\frac{m_1}{\sqrt{m_0+m_2}},
\qquad 
  A_{22}=\frac{3}{4}k\frac{\tilde{a}_1^2}{\tilde{a}_2^{7/2}}\frac{m_1}{\sqrt{m_0+m_2}}, \feq
where $\tilde{a}_1$ and $\tilde{a}_2$ are the averaged semimajor axes measured 
from $m_0$ to $m_1$ and $m_2$, respectively.

Now we rewrite the system in terms of familiar quantities to find
\beq \frac{de_1}{dt}=A_{12}e_2\sin g, \label{eq:de1dtcl} \feq
\beq \frac{de_2}{dt}=-A_{21}e_1\sin g, \feq
\beq \frac{dg}{dt}=A_{22}-A_{11}+A_{12}\frac{e_2}{e_1}\cos g-
A_{21}\frac{e_1}{e_2} \cos g, \feq
where $g=g_2-g_1$. Indeed the use of this variable is convenient, since for
coplanar orbits there is no well-defined line of nodes, and only the 
relative longitudes of perihelia are important.

Upon expanding the octupole equations~(\ref{eq:oct1a})--(\ref{eq:oct1b})
to first order
in $e_1$ and $e_2$ we obtain an identical linear system
of differential equations, but with the $A$'s replaced by
\beq
 B_{11}&=&\frac{12C_2}{L_1}= \frac{3}{4}k\frac{a_1^{3/2}}{a_2^3}
  \frac{m_2}{\sqrt{m_0+m_1}}, \\
 B_{12}&=&\frac{4C_3}{L1} \ = \frac{15}{16}k\frac{a_1^{5/2}}{a_2^4}
  \frac{m_2(m_0-m_1)}{(m_0+m_1)^{3/2}}, \\
 B_{21}&=&\frac{4C_3}{L_2}=\frac{15}{16}k\frac{a_1^3}{a_2^{9/2}}
  \frac{m_0m_1(m_0-m_1)\sqrt{m_0+m_1+m_2}}{(m_0+m_2)^3}, \\
 B_{22}&=&\frac{12C_2}{L_2}=\frac{3}{4}k\frac{a_1^2}{a_2^{7/2}}
  \frac{m_0m_1\sqrt{m_0+m_1+m_2}}{(m_0+m_1)^2}.
\feq
It is easy to see that, in the
limit where $m_0 \gg m_1$ and $m_0 \gg m_2$, the two sets of
equations coincide, as they should. This establishes the accuracy
of our analytic results in this limit.

In general, the two sets of equations differ in the dependence of 
the coefficients on the masses. Although the two theories use 
different coordinate systems (Jacobi vs heliocentric), this alone 
does not explain the difference. Instead, one must remember
that the classical theory was derived from Lagrange's planetary
equations (see Brouwer \& Clemence 1961), which assume that the
disturbing functions (proportional to $m_1$ and $m_2$) are
small. Therefore the approximation is valid only if $m_0 \gg m_1,m_2$.
This can also be seen by considering the
$m_0=m_1$ case, for which Heggie \& Rasio (1996, App.~B) proved that
the variation of $e_1$ vanishes to all orders in $\alpha$ if the
initial $e_1=0$. In contrast, the classical planetary perturbation 
equation~(\ref{eq:de1dtcl}) 
would predict a nonzero perturbation of $e_1$ for this case
(since $A_{12}\ne 0$, while our coefficient $B_{12}=0$ for $m_0=m_1$).

Our octupole-level analytic results do not depend on any assumption
about the three masses, as long as the system can be modeled as a
hierarchical triple.  
The octupole equations predict constant eccentricities in the case
where $m_0=m_1$ and $i=0$. This happens because the odd-power
coefficients are proportional to $m_0-m_1$ in the Hamiltonian~(\ref{eq:Ham1}), 
so the leading terms vanish.  There is no reason to
expect the octupole approximation to work for this very particular case.
However, in \S 3.6 we show explicitly by comparison with direct
numerical integrations that the mass-dependences of our 
equations~(\ref{eq:oct1a})--(\ref{eq:oct1b}) are valid for 
wide ranges of both mass ratios.


\section{Comparison with Direct Numerical Integrations}

We have performed extensive numerical integrations of hierarchical
triple systems using both our octupole-order secular perturbation
equations (hereafter OSPE) and direct three-body integrations.  In
this section we present a sample of results that establish the
validity and accuracy of our analytic results, and at the same time
illustrate the dependences of the perturbations on different
parameters.

\subsection{Numerical Methods}

For the numerical integration of the OSPE we change variables from
($e_1$, $g_1$, $e_2$, $g_2$) to ($e_1 \sin g_1$, $e_1 \cos g_1$, $e_2
\sin g_2$, $e_2 \cos g_2$) to remove singularities associated with the
longitude of pericenter for circular orbits.  We perform the numerical
integration of the OSPE using the
Burlisch-Stoer integrator of Press \etal (1992).  Energy and angular
momentum are automatically conserved, since the semimajor axes are
considered constant and the relative inclination of the orbits varies
to conserve angular momentum.  We present results obtained using an
accuracy parameter ${\rm EPS}=10^{-8}$.  We found that reducing the
integration step sizes did not make a significant numerical difference
for several test systems.

We have compared the results of the OSPE integrations with 
direct three-body integrations using a fixed timestep
mixed-variable symplectic (hereafter MVS) integrator (Wisdom \& Holman 1991)
available in the software package SWIFT (Levison \& Duncan 1991).  For
most integrations, we used a timestep of $P_1/40$, where $P_1$ is the
orbital period of the inner binary.  Energy and angular momentum were
typically conserved to one part in $10^6$ and $10^{12}$, respectively.
For some high-eccentricity systems 
we reduced the timestep to a value as small as
$P_1/600$.  This MVS integrator was designed for systems in which
$m_0$ is much larger than both $m_1$ and $m_2$.  For systems with $m_1/m_0
\go 0.1$ we made use of a newer MVS integrator modified to accommodate 
arbitrary mass ratios, kindly provided to us by Jack Wisdom and Matt Holman.
In a number of test calculations with this newer integrator, we varied the 
timestep systematically to verify that our results are not affected 
by numerical errors.

To complement integrations using the MVS integrator, we also used
SWIFT's Burlisch-Stoer (BS) integrator in a few test runs.  This integrator
is valid for arbitrarily strong interactions between any pair of bodies, but is
not well suited for very long integrations.  We have
only performed a small number of these tests, since the BS
integrations require a much longer computer time.  Energy and angular
momentum in BS integrations were typically conserved to one part in
$10^5$.  Most BS integrations were stopped after one full oscillation of
$e_1$, which we used to determine the ``maximum eccentricity
perturbation'' of the inner orbit, although, as pointed out in \S 2.3,
the true long-term secular evolution of the eccentricity will not, in
general, be strictly periodic.  A typical BS run lasting for $\sim 10^5
P_2$ took about 400 CPU hours on a MIPS R10000 processor, while the
same run using the MVS integrator would only take about 2 CPU hours.

The numerical integrations all started with initial values of the
inner and outer arguments of pericenter of $0^{\circ}$ and
$180^{\circ}$, respectively.  This choice leads
to the maximum eccentricity induced in the inner binary in both the
planetary limit (see \S 2.4)  and the quadrupole approximation
(see \S 2.3).  We have performed additional integrations to
verify that the remaining angles (longitudes of ascending node and
initial anomalies) do not significantly affect the secular evolution
of the system.  For large inclinations, the initial argument of pericenter of 
the inner
binary is important in determining whether the system will undergo circulation or 
libration, if the inner orbit has a significant initial eccentricity (see Fig.~3).  
However, if the inner orbit is nearly circular initially, then the initial
values of the angles are of little importance
since the inner orbit can switch from circulation to libration and vice versa. 
For coplanar orbits, the magnitude of the angular momentum of the inner 
orbit increases when
$g=g_2-g_1 <0$ and decreases when $g>0$.  For small, nonzero inclinations this
can still serve as a guide when considering the effect of varying the angles.

\subsection{Eccentricity Oscillations}

\begin{figure}
\plotone{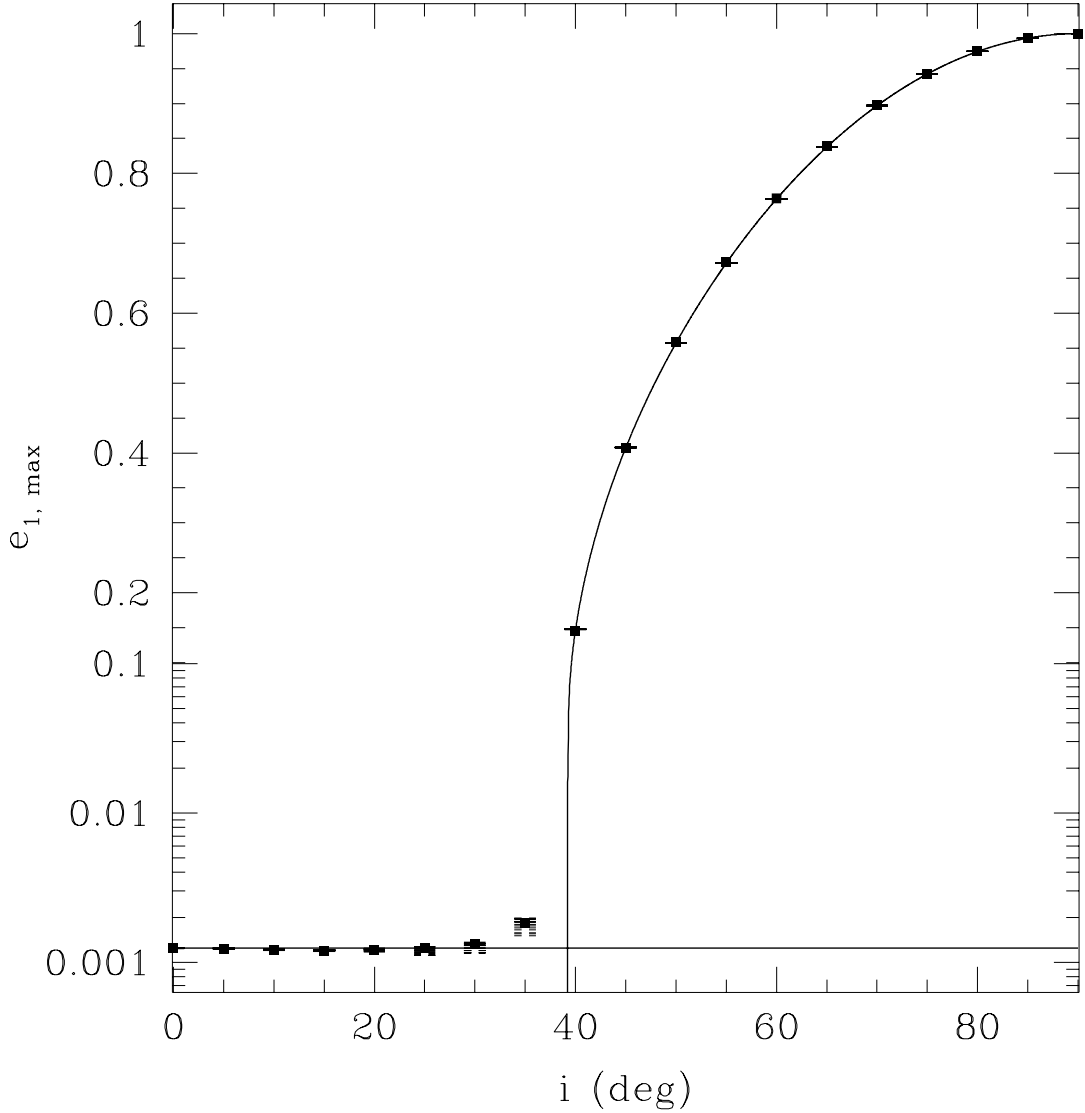}
\caption[Maximum Eccentricity versus Inclination]
{Maximum eccentricity of the inner orbit after a single oscillation, as
a function of the relative inclination.  Here $m_1/m_0 = 10^{-3}$,
$m_2/m_0 = 0.01$, $\alpha^{-1} = 100$, $e_2 = 0.05$, and the initial $e_1 =
10^{-5}$.  The squares are from MVS integrations, and the
double dashes on either side are from OSPE integrations with varying 
initial longitude of
periastron.  The horizontal line indicates the amplitude of the
eccentricity oscillations calculated analytically in the planetary
theory (\S 2.4).  The solid curve indicates the amplitude of
eccentricity oscillations calculated analytically according to the
quadrupole-level theory for $i\go 40^{\circ}$.}
\end{figure}

First, we investigate the dependence of the maximum eccentricity
perturbation of the inner orbit on the initial relative
inclination (see Fig.\ 6).  We see that, as expected (Sec.~2.3), for
small inclinations, $i\lo 40^{\circ}$, the perturbations are dominated
by the octupole term, while for higher inclinations the quadrupole-level
perturbations dominate. In {\it both\/} regimes the OSPE results match 
the direct numerical integrations very well.  Near the transition,
numerical integrations (both OSPE and MVS) show a beat-like pattern of
eccentricity oscillations suggesting an interference between the
quadrupole and octupole terms.  
Note that the results of Figure~6 are for a system with $m_1 \ll m_0$
and $m_2 \ll m_0$, for which the analytic results from the classical
planetary perturbation theory (Sec.~2.4) can be applied for small
eccentricities and inclinations. We see that the agreement with both MVS 
and OSPE integrations is excellent for $i\lo 30^{\circ}$.

We now discuss in some more detail the evolution of systems in the
low- and high-inclination regimes.

\subsubsection{Large-inclination Regime}

Figure~7 illustrates the evolution of
$e_1$, $g_1$, $e_2$, $g_2$, and $i$ obtained from a numerical
MVS integration for a typical system with large relative inclination.  
For large inclination, the secular quadrupole-level perturbations
dominate the evolution.  In the quadrupole approximation the inner
eccentricity undergoes periodic oscillations, while the outer
eccentricity remains constant. Indeed, we see in Figure~7 that
$e_1$ undergoes approximately periodic oscillations of large amplitude
(with corresponding oscillations in $i$),
while $e_2$ remains approximately constant. The small-amplitude
(about 10\%) fluctuations in $e_2$ are due mainly to the smaller, 
octupole-level perturbations.

Deviations from strict periodicity in the variation of $e_1$ and $i$
are also caused by octupole-level perturbations.
The period of a quadrupole eccentricity
oscillation is a function of the mass ratios and the outer
eccentricity (see \S 2.3).  Our numerical integrations of
the OSPE reveal that the most significant corrections to this period
come from the variable time spent at low eccentricities.
Equation~(\ref{eq:quad1b}) implies that the time derivative of $e_1$
is small when the inner orbit has a small eccentricity.  Then the
octupole (and higher-order) perturbations can become important,
causing significant variation in the time a system will spend with a
small inner eccentricity. This effect can be seen clearly in Figure~7.

The OSPE do not correctly describe the evolution of a
systems starting with $e_1=0$.  However for any system with
arbitrarily small but nonzero $e_1$, the inner orbit can switch back
and forth between libration and circulation in the $e_1$-$g_1$ plane,
achieving the full range of eccentricities.  In our MVS integrations
the short-period variations (averaged out in secular perturbation theory) 
provide the necessary perturbations to allow
for the full eccentricity oscillations, even if the initial $e_1=0$.

\begin{figure}
\plotone{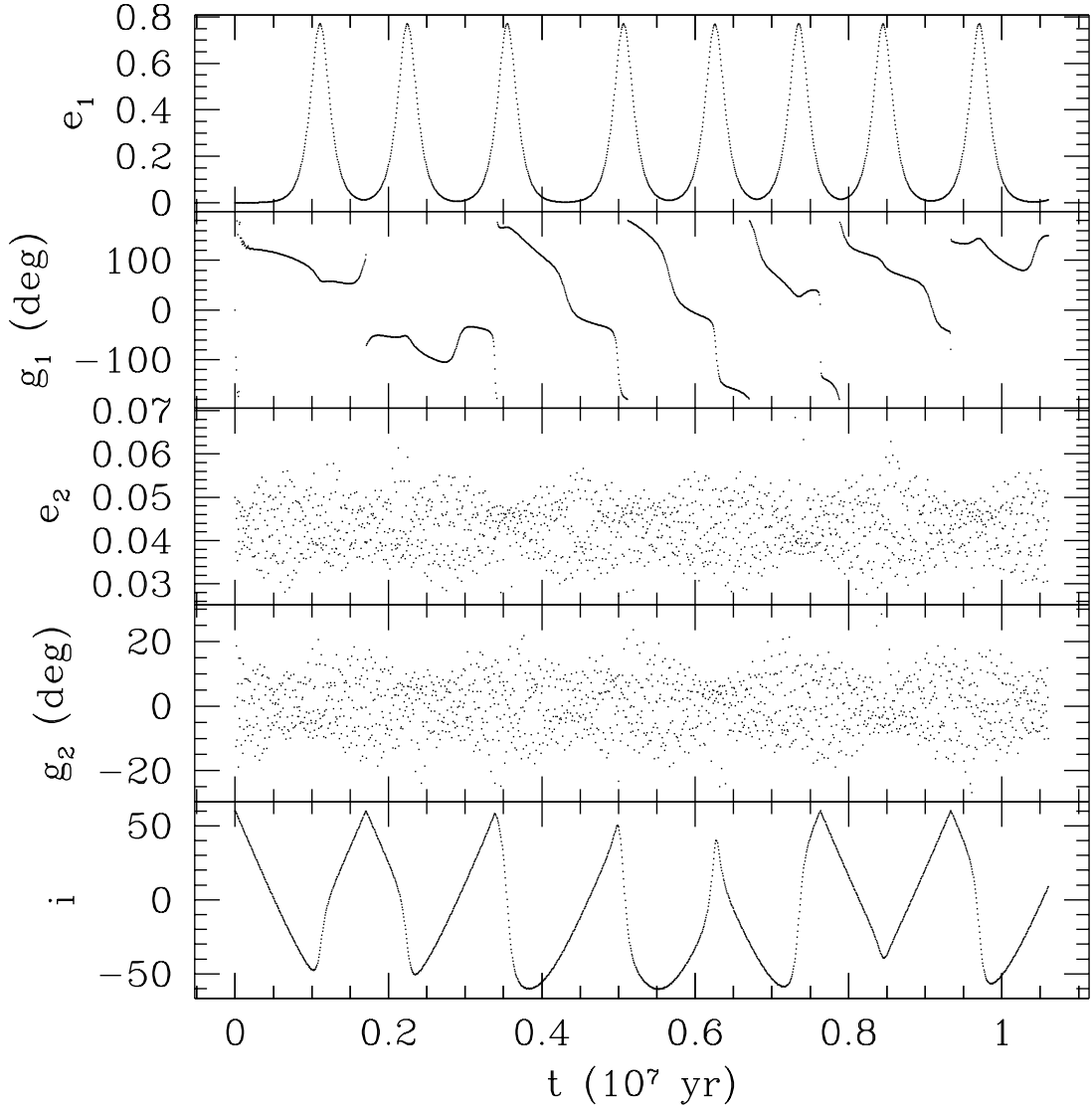}
\caption[Secular Evolution of a System with Large Relative Inclination]
{Typical evolution of the eccentricities, longitudes of periastron, and 
relative inclination for a
system in the high-inclination regime. Here $m_1 / m_0 = 10^{-3}$, 
$m_2/m_0 = 0.01$, $\alpha^{-1} = 100$, the initial inclination
$i=60^{\circ}$, and the initial eccentricities $e_1 = 10^{-5}$
and $e_2 = 0.05$. 
Time is given in years assuming $a_1=1\,$AU and $m_0=1\,M_\odot$.
These results were obtained using numerical MVS integrations.}
\end{figure}

\subsubsection{Small-inclination Regime}

For small inclinations ($i\lo 40^{\circ}$), the secular octupole-level
perturbations dominate and both $e_1$ and $e_2$ typically undergo
very small-amplitude fluctuations, as does the relative inclination
(Fig.~8).  In the octupole approximation angular
momentum is periodically transferred from one orbit to the other.  In
this regime, the special case of initially circular orbits is stable
to eccentricity oscillations.  Short-period perturbations will still
cause small fluctuations in both eccentricities, but since the
inclination does not undergo large oscillations, the angular momentum
transferred from one orbit to another is limited by angular momentum
conservation, preventing large eccentricities from developing in
either orbit.

\begin{figure}
\plotone{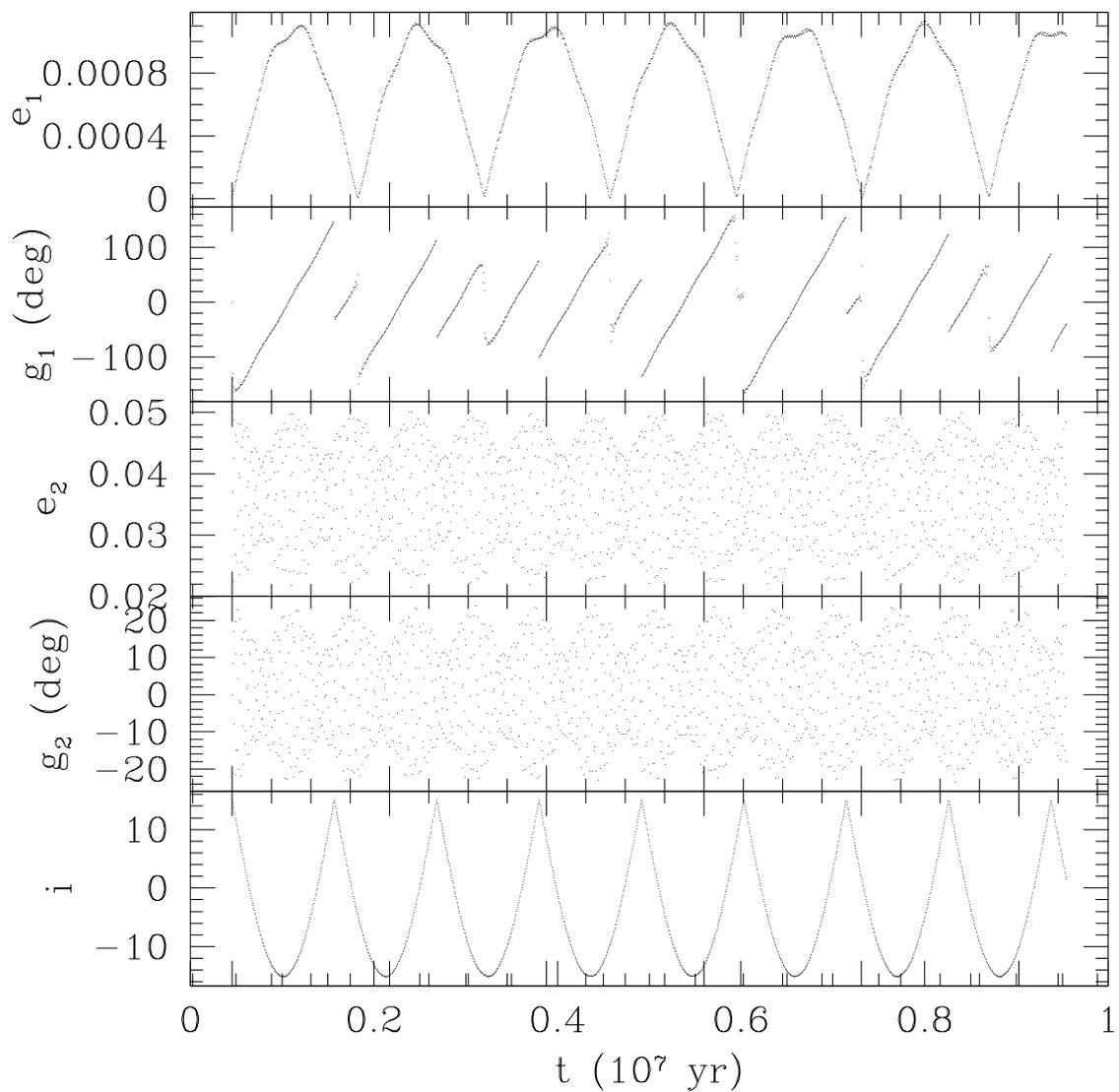} 
\caption[Secular Evolution of a System with Small Relative Inclination] 
{Typical evolution of a system in the low-inclination regime.
All parameters are as in Fig.~7, except that the initial inclination
$i=15^{\circ}$. These results were obtained using numerical MVS
integrations.}
\end{figure}

 For inclinations approaching $\sim 40^{\circ}$, the octupole-level 
interaction still leads to a 
 noticeable amplitude oscillation superimposed onto 
the quadrupole-level result.  For a small range of inclinations the 
two eccentricity oscillations can become comparable leading to 
a secular evolution with a period and amplitude larger than either 
of the two oscillations in isolation.  

\subsection{Dependence on the Ratio of Semimajor Axes}

Numerical results illustrating the dependence of the maximum
eccentricity perturbation of the inner orbit on the ratio of semimajor
axes are shown in Figure~9.  Some small deviations between the OSPE and MVS
results appear for small $a_2/a_1$, where higher-order secular
perturbations may be significant. 
As predicted by the quadrupole-level approximation, the
amplitude of the eccentricity oscillations becomes independent of
$\alpha$ for high inclinations.

\begin{figure}
\plotone{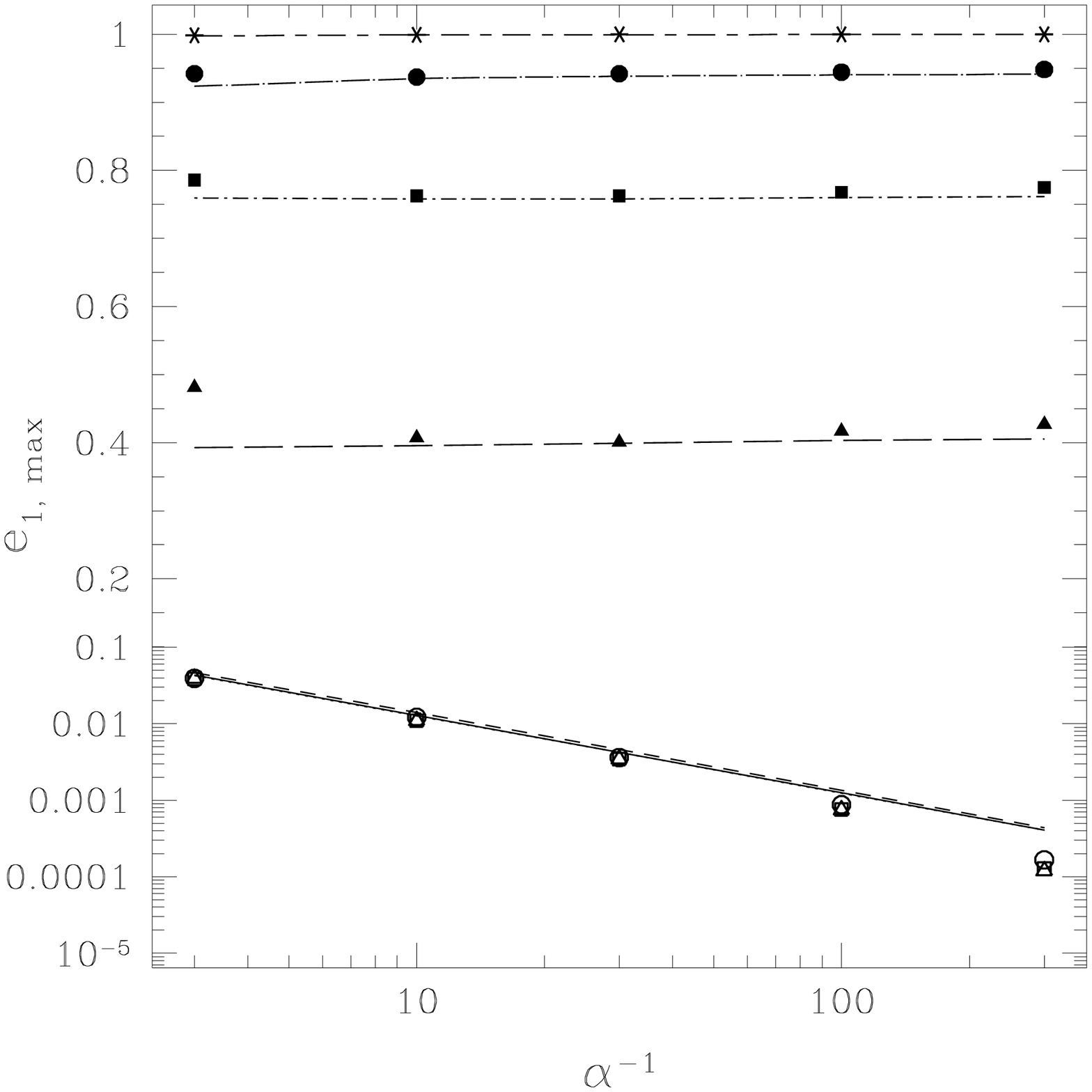}
\caption[Maximum Eccentricity versus the Ratio of Semi-major Axes]
{Maximum eccentricity $e_1$ as a function of the ratio of semimajor
axes $\alpha^{-1}=a_2/a_1$.  The integrations are for a system with $m_1/m_0 =
10^{-3}$, $m_2/m_0 = 0.01$, and initial eccentricities
$e_1=10^{-5}$ and $e_2 = 0.05$.  The various symbols (lines) are from MVS (OSPE)
integrations of systems with various relative inclinations:
$0^{\circ}$ (open triangles, solid line), $15^{\circ}$ (open
squares, dotted line -- here quasi-indistinguishable from the solid
line, cf.\ subsequent figures), $30^{\circ}$ (open circles, short-dashed
line), $45^{\circ}$ (solid triangles, long-dashed line), $60^{\circ}$
(solid squares, short-dash-dotted line), $75^{\circ}$ (solid circles,
long-dash-dotted line), and $89^{\circ}$ (stars, short-dash-long-dashed 
line). }
\end{figure}

\subsection{Dependence on the Initial Eccentricity}

Figure 10 shows the dependence of the maximum inner eccentricity on
its initial value.  For low inclinations increasing the inner eccentricity 
nearly adds to the
maximum induced eccentricity.  In the high-inclination regime
increasing the initial inner eccentricity does not affect the maximum
inner eccentricity significantly until the two become comparable.

\begin{figure}
\plotone{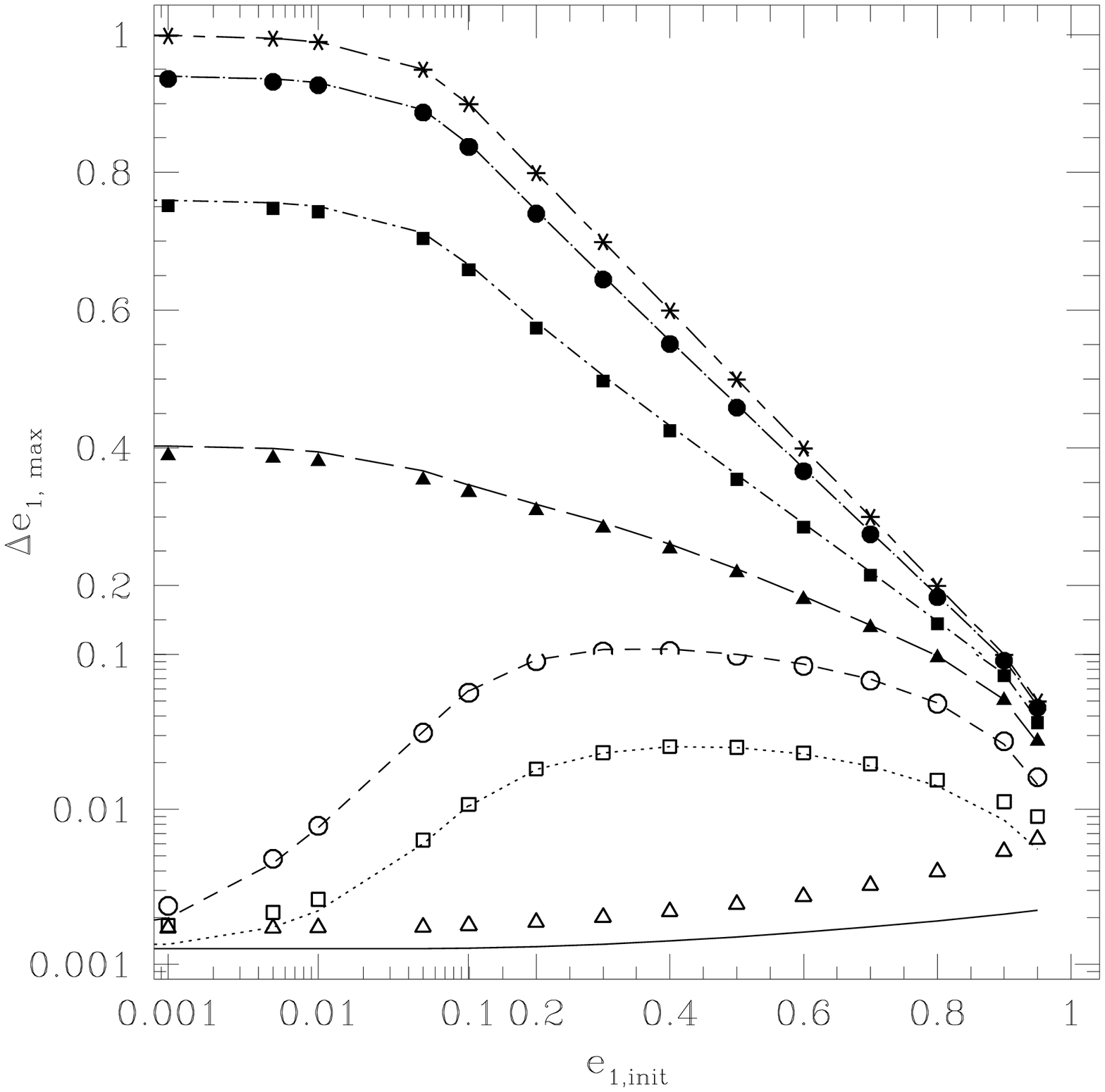}
\caption[Maximum Eccentricity versus the Inner Eccentricity]
{Maximum change in $e_1$ as a function of its initial value.
These integrations are for $m_1/m_0 = 10^{-3}$, $m_2/m_0
= 0.01$, $e_2 = 0.05$, and $\alpha^{-1}=100$.  The symbols and lines
are as in Fig.~9.}
\end{figure}

\subsection{Dependence on the Outer Eccentricity}

Figure~11 shows the effect of varying the outer eccentricity.  The
OSPE and MVS integrations agree precisely for moderate eccentricities,
but show discrepancies for both very large and very small $e_2$.  

For low $i$ and $e_2 \lo 10^{-2}$, short-period eccentricity variations 
become important.  These are not included in the OSPE since they were 
averaged out of the Hamiltonian.  Formally, $i=e_1=e_2=0$ is a fixed point, 
since it implies $de_1/dt$=0 in the OSPE.  For low inclinations
and eccentricities, the short-period eccentricity oscillations determine
the maximum eccentricity, since this fixed point is stable to the
small-amplitude short-period perturbations.  However, for large
relative inclinations, the initial condition $e_1 \simeq e_2\simeq 0$ will lead
to large amplitude oscillations as discussed in \S3.2.1.
Thus, the small-amplitude, short-period oscillations not included in
the OSPE allow the system to explore the full range of allowed
eccentricities.

For large $e_2$, some discrepancies may be
caused by inaccuracies in the MVS integrations: the fixed timestep
implies that periastron passages may not be fully resolved.  We
have performed additional MVS integrations with a smaller timestep
(shown in Fig.~11) and a smaller number of BS integrations to verify
that most of the discrepancy is indeed caused by inaccuracies in the
MVS integrator and not the OSPE.  However, for sufficiently large
$e_2$, the disagreement remains.  As periastron passages begin to resemble 
close dynamical encounters, the averaging over orbits becomes
invalid, and the OSPE are no longer applicable.  In this limit where
the outer orbit is nearly parabolic, it may be better to treat each
periastron passage as a separate encounter. The results of Heggie \&
Rasio (1996) may be used to calculate analytically the eccentricity
perturbation of the inner binary after each encounter.

\begin{figure}
\plotone{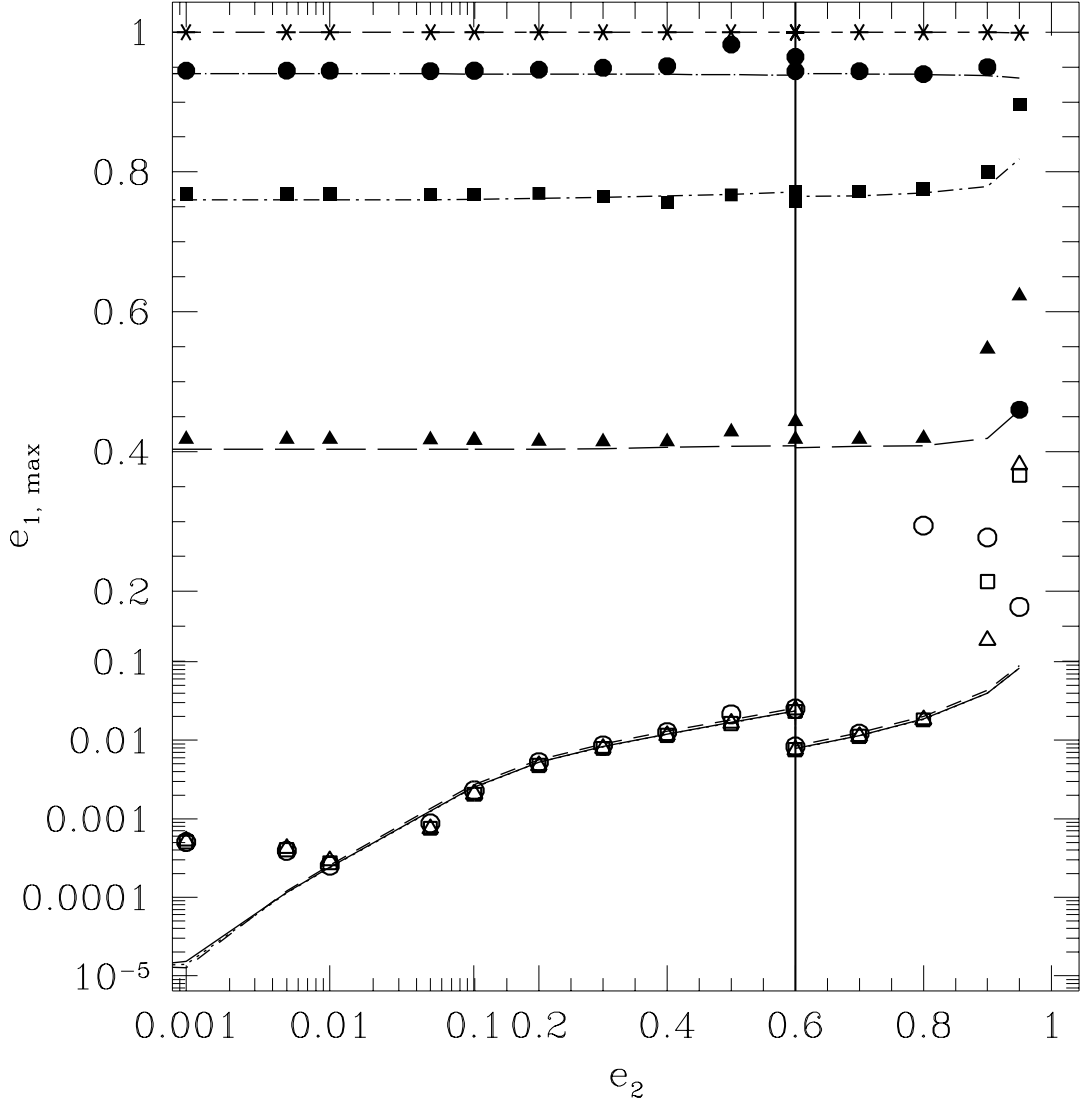}
\caption[Maximum Eccentricity versus the Outer Eccentricity]
{Maximum $e_1$ as a function of the
initial outer eccentricity, $e_2$, for a system with
$m_1/m_0 = 10^{-3}$, $m_2/m_0
= 0.01$, and initial $e_1=10^{-5}$.  For $e_2 < 0.6$ we used $\alpha^{-1}=100$ 
as in previous figures, while 
for $e_2 > 0.6$ we increased the value to $\alpha^{-1} = 300$ 
(to avoid close interactions with $m_2$).  The symbols and line styles are
as in Fig.~9.}
\end{figure}

\subsection{Dependence on the Mass Ratios}

First, we investigate the dependence of the maximum
induced $e_1$ on $m_2$ (Fig.\ 12).  
The MVS and OSPE integrations are in excellent agreement,
 except for very large $m_2$.  For sufficiently large
$m_2$, the binding energy of $m_1$ to $m_2$ becomes comparable to its
binding energy to $m_0$, and the inner orbit deviates significantly from
a Keplerian orbit, making the basic assumption of a hierarchical triple
invalid.  As discussed in \S 3, the MVS integrator was not
designed for large $m_2/m_0$.  However, we have performed a number of
test integrations, both BS and MVS (with a smaller timestep), and found
that the MVS integrations are generally accurate even for 
$m_2 \go m_0$, provided that the system is stable.

\begin{figure}
\plotone{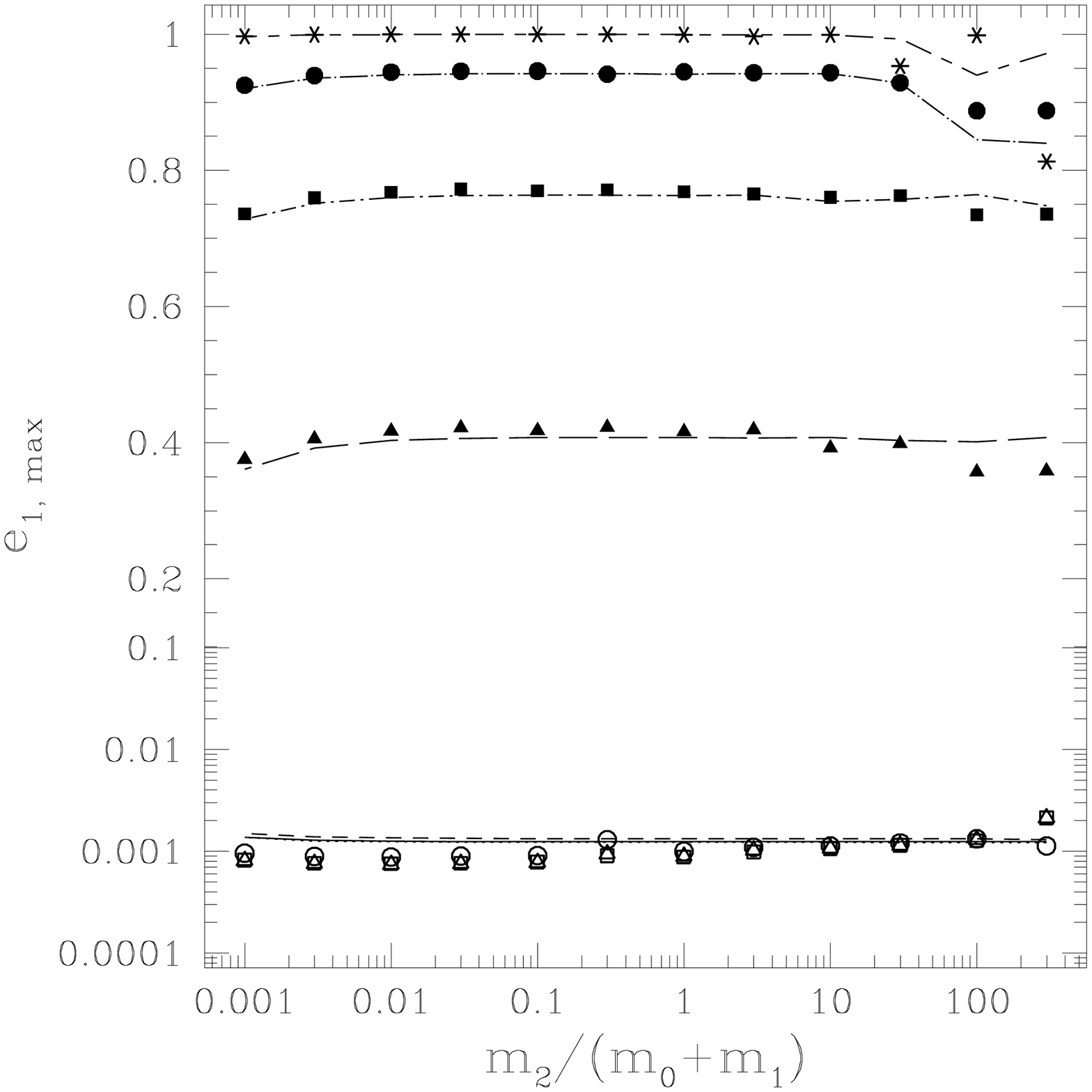}
\caption[Maximum Eccentricity versus the Outer Mass]
{Maximum $e_1$ as a function of the mass
of the outer body, $m_2$, for a system with
$m_1/m_0 = 10^{-3}$, $\alpha^{-1} = 100$, and initial eccentricities 
$e_1 = 10^{-5}$ and $e_2 = 0.05$.  The symbols and line styles are as in Fig.~9.}
\end{figure}

Next, we explore the effects of varying the ratio $m_1/(m_0+m_1)$
(Fig.\ 13). The agreement between MVS and OSPE results is very good, even
when $m_1\simeq m_0$.
In the OSPE, the octupole-level
perturbations vanish when $m_2/m_1 = 1$, removing the dominant term of the
expansion for low inclinations.  Therefore we did not expect the OSPE
to properly model the systems with low inclinations.  
Using the modified MVS integrator of Wisdom
and Holman (see \S3.1) for $m_1/m_0 \go 0.1$, 
we find surprisingly good agreement between
the OSPE and MVS results for both low and high inclinations.  In
particular the OSPE and MVS integrations agree on the maximum
induced eccentricity in the equal-mass case, $m_1= m_0$, which is an
important case for binary stars.  Additionally, the OSPE and
MVS results show similar peaks in the maximum induced eccentricity around the
resonance between the quadrupole and octupole terms. Both MVS and limited 
BS integrations also indicate that the vanishing of the
induced eccentricity for low-inclination systems when $m_0 = m_1$ is real. 
(Unfortunately,
these systems are very time-consuming to integrate numerically with a
BS integrator, prohibiting us from doing a more thorough investigation.)
For example, for the initial
conditions $m_0=m_1=m_2$, $e_1=e_2=0$, $\alpha^{-1}=100$, and $i=0$, we
observed only short-term eccentricity fluctuations of magnitude $\sim
10^{-12}$.  Thus, we conclude that the OSPE results are accurate for all values 
of the inner mass ratio $m_1/(m_0+m_1)$. 

\begin{figure}
\plotone{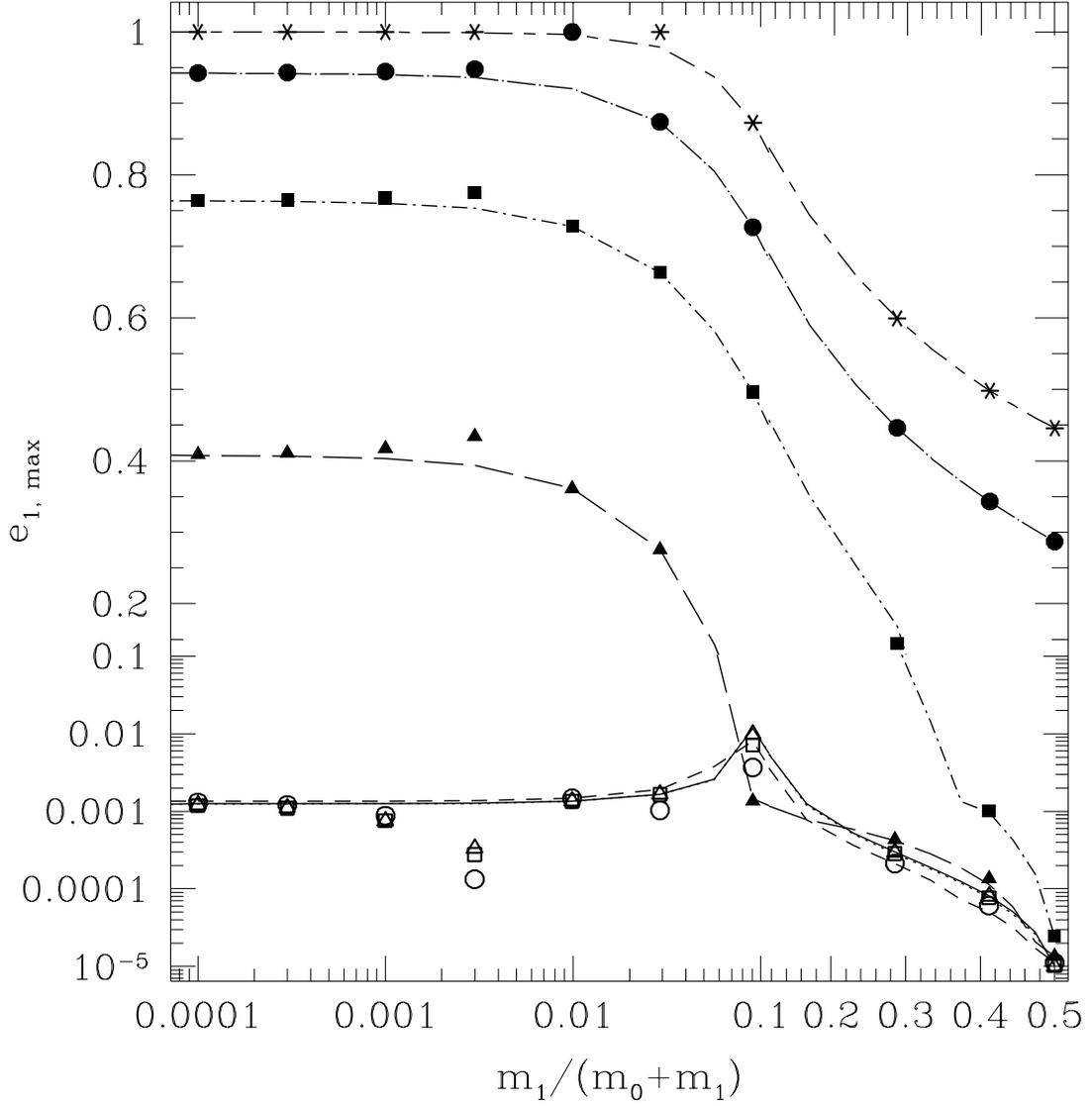}
\caption[Maximum Eccentricity versus the Inner Mass]
{Maximum $e_1$ as a function of the mass ratio
of the inner binary, $m_1/(m_0+m_1)$, for a system with
the same parameters as in Fig.~12.
The symbols and line styles are as in Fig.~9.}
\end{figure}

\subsection{Summary and Discussion}

We have performed a large number of numerical integrations (including
many not shown here) to establish the validity of our analytic results
for a broad range of triple configurations.  
The only significant difference we observed was in the regime where $e_1\simeq
0$. In that regime the system will chaotically choose circulation or libration
about an island in the $(e_1, g_1)$ phase space.  Since $e_1=0$
creates a singularity in the OSPE, we circumvented this problem by
starting runs with $e_1=10^{-5}$.  While varying the timestep affected
when $m_1$ chose to librate or circulate, it did not create any
significant difference in the ratio of circulation to libration time.

We conclude that the OSPE provide an accurate description of the
secular evolution of hierarchical triple systems (containing
unevolving point masses and in Newtonian gravity) for nearly all
inclinations, initial eccentricities, and mass ratios.  The OSPE may
be used for small $e_1$, provided that $e_1 \ne 0$, since this can be 
unstable to large oscillations.  When secular perturbations are
sufficiently small, short-period perturbations may provide the larger
contribution to the eccentricity oscillations.  The OSPE are not
applicable when $m_0/m_2 < \alpha \equiv a_1/a_2 $, since the inner orbit
is then no longer nearly-Keplerian.  The OSPE also break down 
whenever $ a_2 (1-e_2) / a_1 (1+e_1) \lo 3-5$, since the triple system 
is then likely unstable and its evolution will not be dominated by secular effects.  
Similarly, the OSPE should not
be applied when $a_1 (1-e_1) \lo R_0$, where $R_0$ is the radius of
the larger of the two inner stars, since the tidal interaction with
that star would then be important.  One should also be careful
whenever $e_1\simeq 1$, since a small fractional error in $e_1$ can lead 
to a significant change in $r_{p,1}\equiv a_1 (1-e_1)$ which is important 
in differentiating purely gravitational interactions from a strongly dissipative 
tidal encounter or collision between the inner components.

\section{Resonant Perturbations: The Case of PSR B1620$-$26}

\subsection{Introduction}

Hierarchical triple systems can be affected by many different types of
perturbations acting on secular timescales.  In general, during a
given phase in the evolution of a triple, only one type of
perturbation will be important. However, it is possible that, in some
cases, two perturbation mechanisms with different physical origins may
be acting simultaneously and combine in a nontrivial manner. In
particular, whenever two perturbations are acting on comparable
timescales, the possibility exists that they will reinforce each other
in a nonlinear way, leading to a kind of resonant amplification. This
is not to be confused with orbital resonances, which can lead to
nonlinear perturbations of two tightly coupled 
Keplerian orbits when the ratio of
orbital periods is close to a ratio of small integers
(see, e.g., Peale 1976)
 
Perturbation effects coming from the stellar evolution of the
components or from tidal dissipation in the inner binary were
mentioned briefly in \S1 and will not be discussed extensively in this
paper. 
Instead, we consider the case where the inner binary contains
compact objects and its orbit is affected by general relativistic
corrections on a timescale comparable to that of the Newtonian secular
perturbations calculated in \S 2. Rather than basing our discussion on
hypothetical cases, we concentrate on the real example provided by the
PSR B1620$-$26 system.

\subsection{The PSR B1620$-$26 Triple System}

PSR B1620$-$26 is a millisecond radio pulsar in a triple system,
located near the core of the globular cluster M4.  The inner binary
consists of a $\simeq 1.4\,M_{\odot}$ neutron star with a $\simeq
0.3\,M_{\odot}$ white-dwarf companion in a 191-day orbit with an
eccentricity of $0.025$.  The mass and orbital parameters of the third
body are less certain, since the duration of the radio observations
covers only a small fraction of the outer period. However, from the
modeling of the pulse frequency derivatives as well as short-term
orbital perturbation effects it appears that the second companion is
most likely a low-mass object ($m_2\simeq0.01\,M_\odot$) in a wide
orbit of semimajor axis $a_2\simeq50\,$AU (orbital period
$P_2\simeq300\,$yr) and eccentricity $e_2\simeq 0.45$ (Joshi \& Rasio
1997; Ford \etal 2000).  The eccentricity of the inner binary,
although small, is several orders of magnitude larger than expected
for a binary millisecond pulsar of this type, raising
the possibility that it may have been produced by long-term secular
perturbations in the triple.

An analysis based on the classical planetary theory (i.e., for small
relative inclination ignoring general relativistic precession) shows that 
a second companion of stellar mass
would be necessary to induce an eccentricity as large as 0.025 in the
inner binary (Rasio 1994, 1995).  Such a large mass for the second
companion has now been ruled out by recent pulsar timing data, and by
the absence of an optical counterpart for the system (Ford \etal
2000). 

It is reasonable to assume 
that the relative inclination is large, since the location of 
the system near the core of a dense globular cluster suggests that the 
triple was formed through a dynamical interaction between binaries
 (Rasio, McMillan, \& Hut 1995; Ford \etal 2000).
For a sufficiently large relative inclination, we have seen (Fig.~6) 
that it should always be possible to induce an arbitrarily
large eccentricity in the inner binary. Therefore, this would seem to
provide a natural explanation for the anomalously high eccentricity of
the binary pulsar in the PSR B1620$-$26 system (Rasio, Ford, \&
Kozinsky 1997).  However, there are two additional conditions that
must be satisfied for this explanation to hold.

First, the timescale for reaching a high eccentricity must be shorter
than the lifetime of the triple system. In this case the lifetime of
the triple is determined by the timescale for encounters with passing
stars in the cluster, since any such encounter is likely to disrupt
the orbit of the (very weakly bound) second companion. As discussed in
detail by Ford \etal (2000), this is unlikely to be the case in
the high-inclination regime of secular perturbations, given the
parameters of PSR B1620$-$26 and its location near the core of M4
(or inside -- it is seen just inside the edge of the core in projection).

Second, the secular perturbation of the inner binary pulsar by its
distant second companion must be the dominant source of orbital
perturbation.  Additional perturbations that alter the longitude of
periastron of the inner binary can indirectly affect the evolution of
its eccentricity.  For a binary pulsar, general relativity contributes a significant
orbital perturbation.  If the
additional precession of periastron induced by general relativity is
much slower than the precession due to the Newtonian secular
perturbations, then the eccentricity oscillations should not be
significantly affected.  However, if the additional precession is
faster than the secular perturbations, then eccentricity oscillations
may be severely damped (Holman \etal 1997; Lin \etal 1998).  
In addition, if the two precession periods are comparable, then a type of
resonance could occur, leading to a significant increase in the
eccentricity perturbation.

\subsection{Secular Evolution of the Eccentricity}

We have used the OSPE to study the secular evolution of the inner
binary eccentricity in the PSR B1620$-$26 system. We integrate the
system using the variables $h_1,\,h_2,\,k_1,\,k_2$ (eqs.\ \ref{eq:h} \&
\ref{eq:k}), which makes it easy to incorporate the first-order post-Newtonian
correction.  We restrict our attention to the one-parameter family of
orbital solutions calculated by Ford \etal (2000), based on the
modeling of the four pulse frequency derivatives measured by 
Thorsett \etal (1999).  For each solution, the
maximum induced eccentricity of the inner orbit depends only on the
(unknown) relative inclination of the two orbits.  In Figure~14, we
show this maximum induced eccentricity as a function of the second
companion's semimajor axis for several inclinations.

\begin{figure}
\plotone{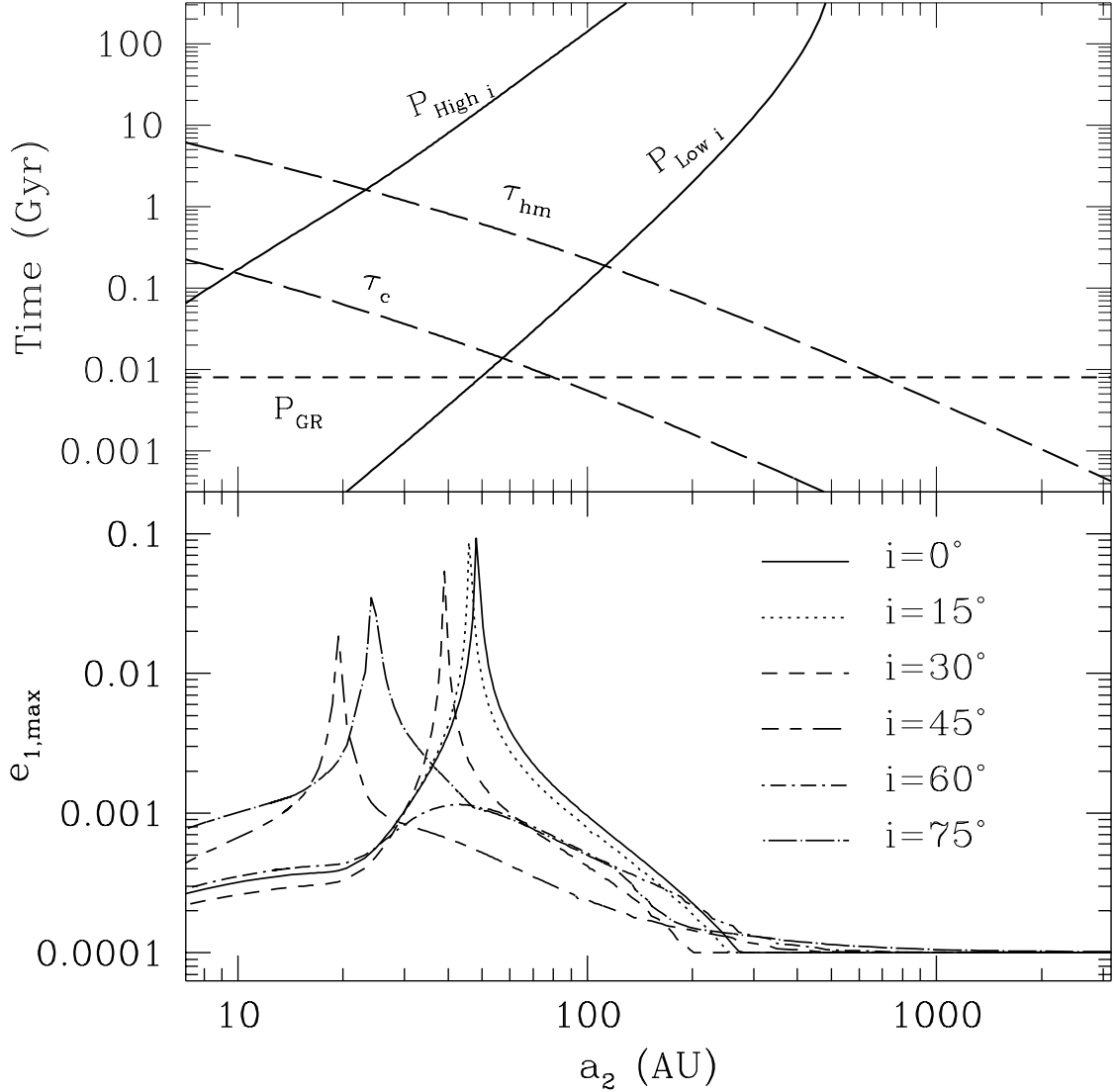}
\caption[Maximum Eccentricity versus Mass for a System with General
Relativistic Precession]
{The top panel compares several timescales in the PSR B1620$-$26
binary pulsar as a function of the semimajor axis, $a_2$, of its
second companion:  $P_{\rm GR}$ is the general relativistic precession
period of the inner binary;  $P_{\rm High~i}$ and $P_{\rm Low~i}$ are
the periods of the eccentricity oscillations in the high and low
relative inclination regimes, respectively; $\tau_{\rm c}$ and
$\tau_{\rm hm}$ are the expected lifetimes of the triple in the
core of M4 and at the half-mass radius, respectively. The bottom panel shows
the maximum induced eccentricity of the inner binary for several
different values of the (unknown) relative inclination.  The peaks
correspond to a resonance between the general relativistic precession
of the inner orbit and the Newtonian secular perturbation by the
second companion.}
\end{figure} 

For most solutions and most values of the inclination, the maximum
induced eccentricity remains significantly smaller than the observed
value of 0.025.  However, for low enough inclinations, there is a
narrow range of solutions (around $a_2\simeq 45$)
for which the observed value can be
reached. Most remarkably, these solutions are also the ones currently
preferred if one takes into account preliminary measurements of the
fifth pulse derivative and short-term orbital perturbation effects in
the theoretical modeling (see Ford \etal 2000).  We also see from
Figure~14 that the maximum induced eccentricity has a peak where the
precession period due to the secular perturbation of the second
companion is comparable to the precession period due to general
relativity, as expected.  As the inclination increases, the maximum
induced eccentricity (at the peak) becomes smaller and the peak moves
towards lower values of $a_2$.  This pattern continues for
inclinations slightly larger the normal cutoff for the low-inclination
limit ($\simeq 40^{\circ}$).  For relative inclinations $50^{\circ}
\lo i \lo 70^{\circ}$ we do not find a peak in the maximum induced
eccentricity.  For inclinations $\go 75^{\circ}$ we again find a peak,
which becomes smaller and moves towards larger separations as the
relative inclination of the two orbits is increased.

The maximum inner eccentricity is also limited in this case by the
relatively short lifetime of the triple system in M4, $\sim
10^7-10^9\,$yr depending on whether the system resides inside or
outside the cluster core (Ford \etal 2000).  For solutions near a
resonance, the inner eccentricity starts growing linearly at nearly
the same rate as it would without general relativistic perturbations.
However the period of the eccentricity oscillations can be many times
the period of the classical eccentricity oscillations.  Although this
allows the eccentricity to grow to a larger value, the timescale for
this growth is also longer. For PSR B1620$-$26, Ford \etal (2000) show
that, near resonance, the inner binary achieves an eccentricity of
0.025 in a time comparable to the expected lifetime of the triple in
the core of M4. However, the location of the pulsar near the edge of the
core (in projection) suggests that it may in fact reside well
outside the cluster core, where its lifetime can be significantly
longer.  In summary, the resonance between general relativistic precession and
Newtonian secular perturbations by the outer companion provides a possible 
explanation for the inner binary's eccentricity.  

\subsection{Comparison with Direct Numerical Integrations}

We have conducted a few long integrations with our MVS symplectic
integrator for model systems similar to PSR B1620$-$26, in order to
check the validity of the OSPE in the presence of a resonance
(Fig.~15).  Although there is good overall agreement, we find that
both the amplitude and the width of the peak is slightly narrower in
the MVS integrations.  Note that the values of the masses and
semimajor axes were decreased in order to speed up the direct
integrations and to satisfy the assumptions of the well-tested MVS
integrator provided in SWIFT (i.e., $m_1\ll m_0$ and $m_2\ll m_0$).
This results in smaller values for the ratio of semimajor axes,
implying less accurate results from the OSPE.  
Nevertheless, the OSPE integrator performs well, even near a resonance 
such as the one
produced by general relativistic precession in a system like
the PSR B1620$-$26 triple.

\begin{figure}
\plotone{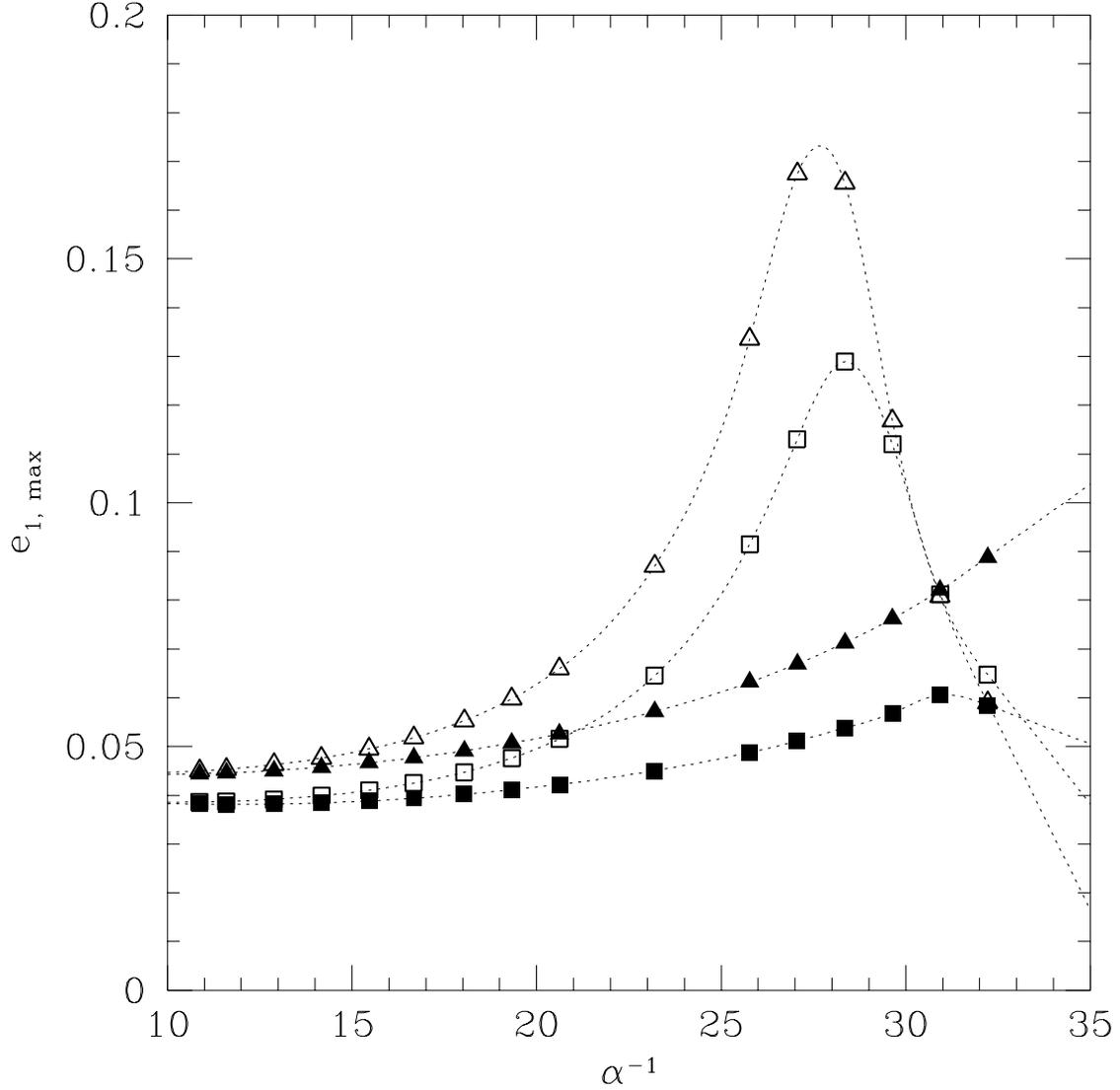}
\caption[GR Resonance Comparison]
{Maximum induced eccentricity $e_1$ as a function of the ratio of
semimajor axes, $\alpha^{-1}=a_2/a_1$. The different symbols show the
results of numerical integrations with and without the general
relativistic term, using both OSPE and MVS integrators: MVS integrations are shown by squares, OSPE integrations are shown by triangles, integrations which include general relativistic precession are indicated by empty symbols, integrations which ignore GR are shown with solid symbols.  These results are for a system similar to PSR B1620$-$26, but with the masses and semimajor axes altered to facilitate the numerical integrations: $m_0=1.4\,M_\odot$, $m_1=5\times 10^{-3}\,M_\odot$,
$m_2=8\times 10^{-4}\,M_\odot$, $e_{1,{\rm init}}=10^{-4}$, $e_{2,{\rm
init}}=0.193$, $a_1= 2 \times 10^{-4}\,$AU, and $i=0^{\circ}$.}
\end{figure} 

\section{Application to Other Observed Triples}

\subsection{The 16 Cygni Binary and its Planet}
 
The 16~Cyg system contains two solar-like main-sequence stars in a
wide orbit (separation $\sim 10^3\,$AU) and a low-mass companion
orbiting 16~Cyg~B in a 2.2-yr ($\sim1.7\,$AU) eccentric orbit ($e=0.67$).
The amplitude of the observed radial velocity variations indicate that
the low-mass companion has a mass $m\sin i \simeq 0.6 M_{\rm Jup}$,
suggesting that it is a giant planet (Cochran \etal 1997).  However
the large orbital eccentricity is surprising for a planet.
 
Holman \etal (1997) and Mazeh \etal (1997) pointed out that the
secular perturbation by 16~Cyg~A could explain the large eccentricity
of the planetary orbit for a sufficiently large relative inclination.
In order for the quadrupole perturbations to be effective the precession of
the longitude of periastron must be dominated by the secular perturbations
of 16~Cyg~A.  The general relativistic precession period ($\sim 7 \times
10^7\,$yr) can be significant in the eccentricity evolution of the planet
(Holman \etal 1997).  
Similarly, if additional companions to 16~Cyg~B are found in larger
orbits (like those recently detected around Upsilon Andromedae; see Butler
\etal 1999), these would also induce a secular change in the longitude
of pericenter of the presently known planet.  Thus, additional companions
could also affect the eccentricity generated by the influence of 16~Cyg~A.  
Such an interaction could prevent the quadrupolar secular
perturbation by 16~Cyg~A from accumulating long enough to produce the
observed large eccentricity.

Hauser \& Marcy (1999) have combined radial velocity and astrometric
data to compute a one-parameter family of solutions, which they
tabulate as a function of the line-of-sight component $z$ of the
position vector of B relative to A. There is
also a possibility that 16~Cyg~A may have an M-dwarf companion which
would affect the orbital solutions for the 16~Cyg~AB binary and hence
the secular perturbation timescale (Hauser \& Marcy 1999).  We have
used their orbital solutions for 16~Cyg~A (with no M dwarf companion)
to estimate the effects of secular perturbations in this system.
If we assume that the eccentricity of 16~Cyg~Bb is due to quadrupolar secular
perturbations, then both general relativity and any additional
planets around 16~Cyg~B could constrain the orbit of the 16~Cyg~AB
binary.  In Figure~16 we compare timescales for eccentricity oscillations
induced by 16~Cyg~A, general relativistic precession, 
as well as the eccentricity
oscillations induced by a hypothetical second planet around 16~Cyg~B.  
While the period of eccentricity oscillations is shorter than $5\,$Gyr
(the approximate age of 16~Cyg~A\&B; see Ford, Rasio, \& Sills 1998, 2000) 
for about $75\%$ of the 
orbital solutions listed by Hauser \& Marcy (1999; we actually used 
an extended version of their Table~4 kindly provided by H.~Hauser), 
the period of the eccentricity oscillations is
shorter than the general relativistic precession period for only
about $25\%$ of their solutions (assuming $\sin i\simeq 1$; with 
$\sin i \simeq 0.5$ this fraction increases to about $60\%$).  

Given the large mass
ratio of 16~Cyg~A to 16~Cyg~Bb and the high eccentricity of the orbit
($>0.53$), the ratio $C_3/C_2$ (see Eq.~\ref{eq:c2} \& \ref{eq:c3}) can approach
unity.  Thus, the octupole term can be very significant in the
dynamics of this system on sufficiently long timescales.
As an example, we use the $z=0$ solution of Hauser \& Marcy (1999) and
assume that the planet was initially on a nearly circular orbit 
with an initial relative inclination of $60^{\circ}$. We find that
the period of eccentricity oscillations is then $\sim 3\times 10^7\,$yr ($\sim
4 \times 10^7\,$yr if general relativistic precession is not included;
$\sim 6 \times 10^7\,$yr if neither general relativistic precession nor the
octupole term is included) and the amplitude of the first eccentricity
oscillation is $\simeq 0.685$ ($\simeq 0.767$ without GR;
$\simeq 0.764$ without GR or octupole term).  However, there is a longer term
oscillation with a period of $\sim7 \times 10^8\,$yr and an amplitude of
0.766 ($\simeq 0.774$ without GR) which is not present when the octupole term
is ignored.  Thus, in this example, the primary effect of general
relativistic precession is to reduce the fraction of the time where the
planet has a very high eccentricity.  
 
\begin{figure}
\plotone{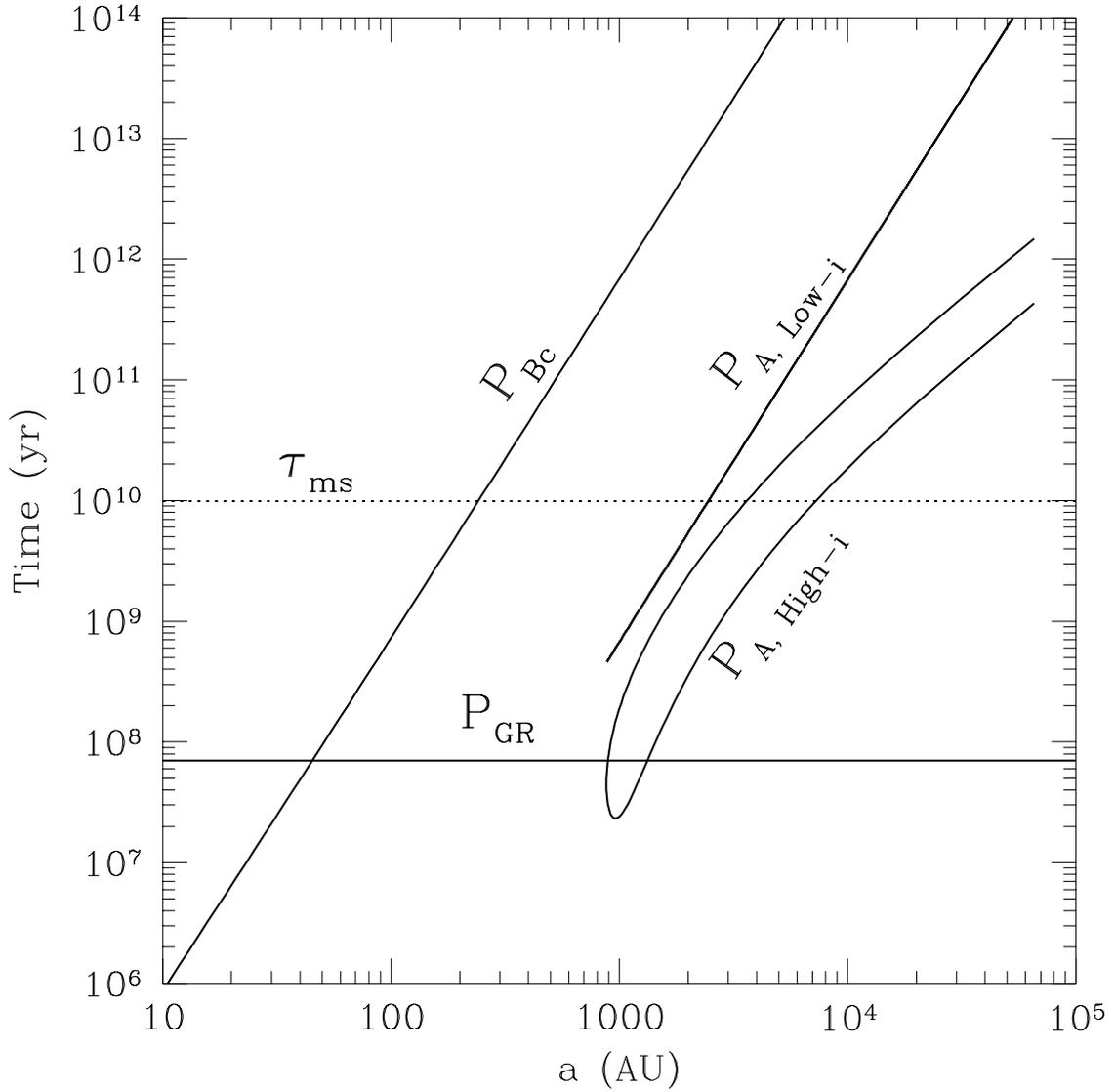}
\caption{This plot compares the timescales for precession of the 
longitude of pericenter of the planet around 16~Cyg~B due to 
the secular perturbations by 16~Cyg~A ($P_{\rm A, High-i}$ and 
$P_{\rm A, Low-i}$ for high and low relative inclination regimes, 
respectively), general relativity ($P_{\rm GR}$), and secular 
perturbations by a hypothetical second planet ($P_{\rm Bc}$), assuming 
that it is coplanar with 16~ Cyg~Bb and has a mass of $1\,M_{\rm Jup}$.
Note that $P_{\rm A, High-i}$, $P_{\rm A, Low-i}$ and $P_{\rm GR}$ 
are plotted as a function of the binary semimajor axis, while
$P_{\rm Bc}$ is shown as a function of the orbital radius of the
additional planet. $P_{\rm A, High-i}$ and $P_{\rm A, Low-i}$ were
calculated for the one-parameter family of orbital solutions given
by Hauser \& Marcy (1999), which do not extend below $a\simeq 900\,$AU.
$\tau_{\rm ms}$ indicates the age of 16~Cyg~B.} 
\end{figure}

\subsection{The Protobinary System TMR-1}
 
HST/NICMOS observations of the TMR-1 system by Terebey \etal (1998)
reveal, in addition to the two protostars (of masses
$\sim0.5\,M_{\odot}$) with a projected separation of 42\,AU, a faint
third body (TMR-1C) that appears to have been recently ejected from
the system. The association of TMR1-C with the protobinary is
suggested by a long, narrow filament that seems to connect the
protobinary to the faint companion.  Given the observed luminosity of
TMR1-C, and using evolutionary models for young, low-mass objects, the
estimated age of $\sim 3\times 10^5\,$yr for the system leads to a
mass estimate of $\sim 2-5 M_{\rm Jup}$. This suggests that the object
may be a planet that was formed in orbit around one of the two
protostars, and later ejected from the system (Terebey \etal 1998).
If the age were increased to $\sim10^{7}\,$yr, the mass would increase
to $\sim 15\,M_{\rm Jup}$, and TMR1-C could also have been a low-mass, 
brown dwarf companion to one of the stars.
 
If TMR-1C is indeed a planet that was ejected from the binary system,
this may place significant constraints on planet formation theory. 
Here we speculate about the process which may have led to the ejection of 
a planet from the TMR-1 system. In the standard planet
formation theory, TMR-1C must have formed in a nearly
circular orbit around one of the protostars.  Secular perturbations by
the other protostar may then have driven a gradual increase in the
eccentricity of the planet's orbit, gradually pushing the system
towards instability.  Large apocentre distances render perturbations
by the other protostar increasingly important.  The planet could then
enter the chaotic regime in which it can
switch into an orbit around the other protostar, possibly switching
between stars many times before finally being ejected from
the system.  The expected velocity after such an ejection is in agreement
with the estimated velocity of TMR-1C (de la Fuente Marcos \& de la
Fuente Marcos 1998).  One concern with this scenario is that the timescale
for ejection may be short compared to the timescale for planet formation.
Early in the evolution the protostellar  disk will damp the planet's
eccentricity.  However, as the planet becomes more massive, the 
gravitational perturbation
by the other protostar becomes dominant.  
In fact, after the protoplanetary core has formed, it may be
able to accrete more mass than in the standard scenario, since it is no longer
confined to accrete from a narrow feeding zone.  
 
We have investigated systems similar to TMR-1,
but with the low mass companion in a nearly circular orbit around one of
the stars.  We assume that the two protostars 
have masses of $1\,M_{\odot}$ and 
$0.5\,M_{\odot}$, with a binary semimajor axis of 50 AU and a 
planet mass of $5\,M_{\rm Jup}$.  
We estimate when the triple system will become unstable by combining our
models of the secular eccentricity evolution of the binary with the
stability criterion of Eggleton \& Kiseleva (1995).  We calculate the
amplitude of secular eccentricity perturbations and see if the system would
violate the Eggleton \& Kiseleva (1995) instability criterion 
($Y\equiv a_2 (1-e_2) / a_1 (1+e_1) <Y_{\rm min}$) when
the planet's orbit reaches its maximum eccentricity.  For orbits with a large
relative inclination, planetary semimajor axes from 14 AU to 8 AU are
expected to become unstable as the relative inclination is varied from
$40^{\circ}$ to $85^{\circ}$.  
 For nearly coplanar orbits, planetary semimajor axes from 16 AU to 3 AU
will become unstable according to this criterion, as the outer eccentricity
is varied from $0$ to $0.8$.  Thus, it seems plausible that a
protoplanet could begin to form near the critical semimajor axis and
eventually be ejected from the system after it has accreted a large amount
of gas.  If we assume the initial semimajor axis of the planet to be 5 AU,
then we can solve for a critical binary eccentricity, which we find to be
0.65.  The period of the eccentricity oscillations responsible for reducing
the instability parameter 
 from $\simeq 10$ to $\simeq 3$ is about $3 \times 10^4\,$yr.
In the coplanar regime we can also apply the stability criteria of
Holman \& Wiegert (1999)\footnote{They define the critical semimajor 
axis as the largest orbital radius for
which planets of all initial longitudes of periastron survived for $10^4$
binary periods.  This is different from the criterion obtained by 
combining secular perturbation theory with the results of
 Eggleton \& Kisseleva (1995). However both criteria provide an 
estimate of when the triple system becomes unstable.  It is 
reassuring that both criteria yield similar results.}.  
As the eccentricity of the TMR-1 binary increases
from $0$ to $0.8$, the critical semimajor axis decreases from 
about $11\,$AU to $2\,$AU. This is precisely
the region where giant planets are expected to form.  If we assume the
initial semimajor axis of the planet to be $5\,$AU, then we can again
solve for a
critical binary eccentricity, which we find to be 0.48.  The two estimates
are in reasonable agreement, and both also agree with the results of
preliminary numerical simulations which we have performed for this system.
 
We have conducted Monte Carlo simulations to study the process of planet
ejection from protobinaries (Fig.~17).  For systems
with large inclinations, the most common outcome for unstable systems is a
collision of the planet with its parent star. However, for
systems with a low relative inclination, the most common outcome was for the
planet to be ejected from the system.  Furthermore, we found that in many
cases it can take up to $\sim 10^7\,$yr for the planet to be ejected.
Since this is longer than a typical planetary formation timescale, the
scenario proposed above appears reasonable.

\begin{figure}
\plotone{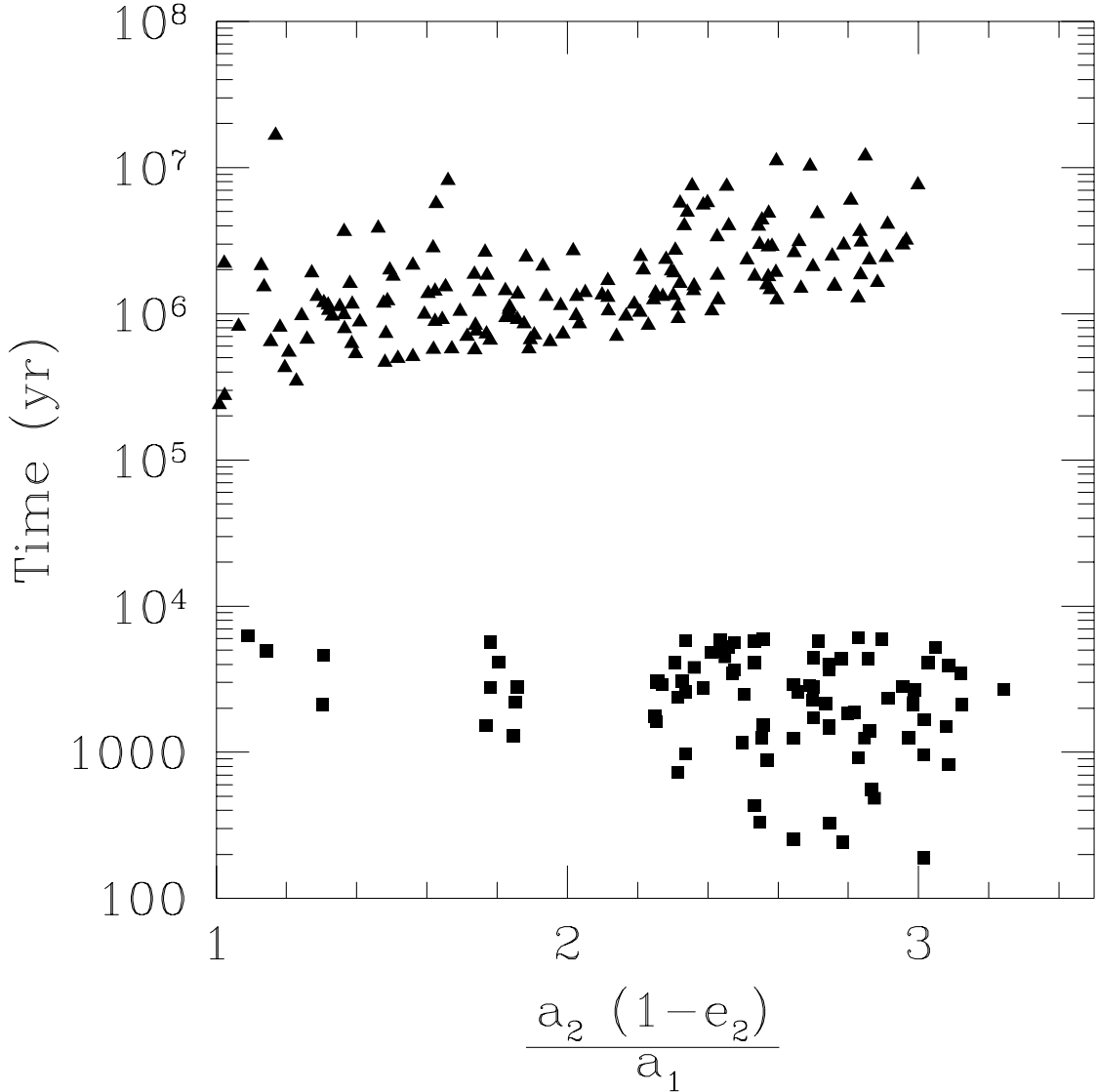}
\caption{
Results of Monte Carlo simulations 
for the dynamical evolution of possible progenitor systems of TMR-1.
The systems contain a
$5\,M_{\rm Jup}$ planet in an initially circular $10\,$AU orbit around a $1
M_{\odot}$ star with a $0.5\,M_{\odot}$ companion star in a $50\,$AU
orbit. All other orbital parameters (initial phases, longitudes of
pericenter, relative inclination, and binary eccentricity)
were assigned random values. 
Triangles correspond to
cases in which the planet was ejected (escaped) and squares correspond
to cases in which the planet collided with one of the stars.
The time to ejection (triangles) or collision (squares) is
shown as a function of the initial stability parameter 
$Y\equiv {a_2 \left(1-e_2\right)}/{a_1}$.  }
\end{figure}

\subsection{Systems with Short-Period Inner Binaries}

Finally, we discuss briefly some observations and related theoretical work
on triple systems containing a short-period inner binary. 
If the outer period is also relatively short,
it may be possible to observe the secular perturbations directly, since the
timescale for eccentricity modulations and orbital precession may become
comparable to the timescale of observations.
Unfortunately, in these systems,
other perturbation effects such as tidal dissipation are likely to affect
the secular evolution, making the theoretical analysis more difficult.

\subsubsection{HD 109648}

HD 109648 is a triple-lined spectroscopic triple probably composed of three
main-sequence stars all with masses $\sim1\,M_{\odot}$
(Jha \etal 1999). The
inner orbit has a short period, $\simeq 5.5\,$d, so tidal dissipation effects 
are likely to be important.  The small but significant eccentricity
($e_1=0.0119\pm 0.0014$) of the inner orbit has been attributed to the 
perturbation by the outer companion (Jha \etal 1999).  This system is strongly 
coupled, with $\alpha\simeq 0.1$, so the timescale for eccentricity 
modulations is short, $P_e\sim 15\,$yr.  Thus, the available
observations, spanning over 8 years, may already have
detected changes in the inner eccentricity and longitude of pericenter.
Theoretical models by Jha \etal (1999) based on current data provide a 
loose constraint on the relative inclination of the orbits: 
$5.9^{\circ} \le i \le 54^{\circ}$ or $126^{\circ}\le i \le 174.1^{\circ}$.
Future observations are likely to produce tighter
constraints on this and other orbital parameters for the triple
system. However, Jha \etal (1999) speculate that additional variations 
may also be caused by the presence of a fourth object in a much wider orbit.

\subsubsection{HD 284163} 

HD 284163 is a triple system in the Hyades.  The inner binary consists
of a $0.72\,M_{\odot}$ primary and a secondary with a minimum mass of
$0.33\,M_{\odot}$ in a 2.4-day orbit (Griffin \& Gunn 1981; Ford \&
Rasio 2000).  The outer companion (of mass $\sim 0.5\,M_{\odot}$) has
a projected separation of $7.4\,$AU (Patience \etal 1998).
Theoretical and empirical evidence indicate that tidal dissipation in
the primary should have circularized the inner binary (Ford \& Rasio
2000).  However, the radial velocity curves indicate a significant
eccentricity, $e_1=0.057\pm 0.005$ (Griffin \& Gunn 1981).  The
secular perturbation by the outer companion is likely responsible for
inducing this observed eccentricity. At present, however, the outer
orbit is not well constrained, making further analysis difficult.
  
\subsubsection{$\beta$ Per} 

This is another triple system with a short-period (2.87 d) inner
binary ($1.7\,M_{\odot}$ + $3.7\,M_{\odot}$) which is expected to have a
very nearly circular orbit on the basis of tidal dissipation theory, but has
a significantly larger observed eccentricity of $0.0653$.  Secular perturbations by
 the
outer companion (mass $1.7\,M_{\odot}$ in a 1.86-yr orbit) are likely responsible for
maintaining the inner binary's eccentricity (Kisseleva \etal 1998).

Kisseleva \etal (1998) suggest that the inner binary may have
originally been significantly wider.  In their scenario, quadrupole
perturbations drive a eccentricity increase.  As the eccentricity
increases, tidal dissipation becomes significant and removes energy
from the orbit.  As the orbit shrinks, precession of the longitude of
periastron due to the stellar quadrupole moments and general relativity
increase, eventually suppressing the eccentricity perturbations.
The secular
decrease in the semimajor axis due to the coupling of quadrupole
perturbations and tidal dissipation is then halted near the presently
observed orbit.

In this system the ratio of semimajor axes is rather small, $\simeq40$.  Thus the
octupole-level perturbations could play an important role in the secular
evolution.  In particular, this could lead to significantly
larger eccentricities in the initial orbit if other effects have not
yet started to suppress the
perturbations.  Thus, the range of initial conditions that could lead
to such an evolution can be much larger than would be expected by
considering quadrupole-level perturbations only.

\acknowledgments

We are grateful to M.~Holman and J.~Wisdom for providing us with a version
of their MVS integrator modified for systems with more than one massive body,
and to H.~Hauser for sending us an extended table of orbital solutions for 
the 16~Cygni binary.
We thank S.~Jha for pointing out to us the recent theoretical
work of Krymolowski and Mazeh and for providing us with a draft of his
paper on HD~109648  in advance of publication.
Some of the numerical simulations mentioned in \S 5.2 for TMR-1 were
performed at MIT by J.~Madic.
F.A.R.\ thanks the Theory Division of the Harvard-Smithsonian
Center for Astrophysics for hospitality. 
This work was supported in part by NSF Grant AST-9618116 and NASA ATP 
Grant NAG5-8460. 
E.B.F.\ was supported in part by the Orloff UROP Fund and the UROP 
program at MIT. 
F.A.R.\ was supported in part by an Alfred P.\ Sloan Research Fellowship.
This work was supported by the National Computational Science Alliance 
under Grant AST980014N and utilized the SGI/Cray Origin2000 supercomputer
at Boston University and the Condor system at the University of Wisconsin.

\end{document}